\documentclass[12pt]{article}

\usepackage[left=1.3in, right=1.3in, top=1in, bottom=1.3in]{geometry}
\usepackage{amsmath, amsthm, amssymb, mathrsfs}
\usepackage[utf8]{inputenc}
\usepackage{rotating} 
\usepackage[comma]{natbib}
\usepackage{ragged2e}
\usepackage[bookmarks=true,pdfauthor=CDtK,colorlinks=true,linkcolor=red,citecolor=blue,urlcolor=red]{hyperref}
\usepackage{longtable}
\usepackage{setspace}
\hypersetup{
	colorlinks=true,
	linkcolor=red,
	filecolor=red,
	urlcolor=red,
	citecolor=red
}

\theoremstyle{definition}

\usepackage{graphicx}
\usepackage[labelsep=space,labelfont=bf,singlelinecheck=on,skip=1pt,sf,justification=centering]{caption}
\usepackage[final]{pdfpages}
\usepackage[capposition=top]{floatrow}
\usepackage{adjustbox}
\usepackage{xcolor}
\usepackage{comment}

\usepackage{pdflscape}

\usepackage{afterpage}

\usepackage{subfigure}
\def\sym#1{\ifmmode^{#1}\else\(^{#1}\)\fi}

\begin{document}

\title{\Large{Global Factors in Non-core Bank Funding and Exchange Rate Flexibility}
\footnote{We thank the Editor, Luc Laeven, and two anonymous referees for very valuable comments on an earlier draft. For comments and discussions, we are grateful to Hilde Bjørnland, Valeriya Dinger, Yiannis Karavias, Samuel Ligonni\`ere, Silvia Miranda-Agrippino, Alessandro Rebucci, Jamel Saadaoui and participants of the Cournot Seminar at the University of Strasbourg, the GREADS Seminar at the University of Birmingham, the 4th Dolomiti Macro Meeting, 2024 Annual Conference of the International Association of Applied Economics and the 2024 International Panel Data Conference. We further thank Daniel Goncalves Koudriachov, Patryk Kuznik and Sanna Wiersinga for outstanding research assistance. An earlier version of this paper was circulated as a CEPR discussion paper DP 18643. Luís Catão gratefully acknowledges the support of a CEPR research fellowship under which much of this research was undertaken. Jan Ditzen acknowledges financial support from Italian Ministry MIUR under the PRIN project %Progetto PRIN 2022 Klimarisiko-Ungewissheit (CRUT) - codice 
2022H2STF2. 
%CUP I53D23002710008, Finanziamento dell'Unione Europea - NextGenerationEU - PNRR M4.C2.1.1. 
Daniel te Kaat gratefully acknowledges financial support from the Dutch Research Council NWO under grant number VI.Veni.211E.023. Additional declarations of interest: none.} 
{\vspace{35pt}} 
}

\bigskip

	\author{Lu\'{i}s A.V. Cat\~ao\footnote{Lisbon School of Economics and Management (ISEG), Research in Economics and Mathematics (REM) and Research Unit on Complexity and Economics (UECE), University of Lisbon (\href{mailto: lcatao@iseg.ulisboa.pt}{lcatao@iseg.ulisboa.pt})} \and Jan Ditzen\footnote{Corresponding author. University of Bozen-Bolzano (\href{mailto:jan.ditzen@unibz.it}{jan.ditzen@unibz.it}) } \and Daniel Marcel te Kaat\footnote{University of Groningen (\href{mailto:d.m.te.kaat@rug.nl}{d.m.te.kaat@rug.nl}) }
 {\vspace{15pt}}}

	\date{\today}
	\maketitle
	\thispagestyle{empty}
\renewcommand{\baselinestretch}{1.0}

\begin{abstract}
We show that fluctuations in the ratio of non-core to core funding in the banking systems of advanced economies are largely driven by three global factors of both real and financial natures, with country-specific factors playing only a minor role. Exchange rate flexibility  helps insulate the non-core to core ratio from such global factors. This insulation is stronger in periods away from global crises. Tighter prudential regulations appear to have a complementary effect to exchange rate insulation.

\bigskip
\bigskip

\noindent {\bf Keywords:}  Global Financial Cycle, Bank Funding, Mundellian Trilemma, Panel Cross-Sectional Dependence, Principal Components, Common Correlated Effects Estimator  		
\end{abstract}
\setcounter{page}{0}
\thispagestyle{empty}
\pagebreak

\renewcommand{\baselinestretch}{1.5}

%\begin{comment}
\newpage
%\tableofcontents
%\newpage
%\listoftables
%\pagebreak
%\listoffigures
%\end{comment}

\pagebreak
\setcounter{page}{1}

\onehalfspacing
%\onehalfspacing

\section{Introduction}
\label{intro}

%cite this one: https://onlinelibrary.wiley.com/doi/abs/10.1111/jmcb.13089?campaign=woletoc

Recent years have witnessed a heated debate in the macroeconomic and international finance literatures on the extent to which global financial conditions influence national ones and, within that, how far capital controls and flexible exchange rates can insulate national economies from global financial shocks. Following the seminal work of \cite{rey2015dilemma}, as well as \cite{borio2014financial}, \cite{bruno2013capital} and others, there has been widespread consensus that global financial conditions---epitomized by the concept of a global financial cycle---are a key determinant of domestic financial cycles and risk assessments by economic agents. 

However, the jury is still out on the extent to which domestic policy instruments can insulate the domestic economy from the global financial cycle. \cite{miranda2020us} and \cite{passari2015financial}, using higher frequency data spanning a wider range of financial instruments over more than 30 years of data, have brought into question the effectiveness of domestic monetary policies and exchange rate flexibility as macroeconomic insulation devices, as advocated by \cite{mundell1963capital} in his famous depiction of the trilemma between exchange rate stability, monetary policy independence and free capital mobility. Yet, several papers, including \cite{aizenman2008assessing}, \cite{aizenman2016monetary}, \cite{obstfeld2019tie}, \cite{obstfeld2015trilemmas} and \cite{shambaugh2004effect}, find evidence that exchange rate flexibility can be effective in insulating domestic interest rates from the global financial cycle, at least partially. In addition, and over and above exchange rate flexibility, various recent studies find that financial account restrictions and some macro-prudential regulations can also be effective insulation devices (e.g., \citealp{akinci2018effective}; \citealp{buch2018cross}; \citealp{cerutti2018changes}; \citealp{fendouglu2017credit}; \citealp{qureshi2011managing}). These latter findings indicate that the Mundellian trillema remains, by and large, a suitable characterization of main trade-offs faced by domestic macroeconomic policy in a globalized world.

In light of ample historical evidence that bank leverage and credit is a key transmitter of business cycle instability (\citealp{shin2012global}; \citealp{brunnermeier2012banks}; \citealp{cecchetti2011weathering}; \citealp{jorda2017bank}), an important extension of this literature has focused on how cross-border banking flows respond to the global financial cycle. \cite{amiti2017supply} find that such flows are well-explained by a common global factor but only during credit expansion periods---in crisis periods, country- and bank-specific factors dominate. This evidence suggests that the importance of the global financial cycle may be time-dependent. A more limited importance of global factors in explaining cross-border financial flows is highlighted in \cite{cerutti2019important}.  Based on a database that covers 85 countries, different types of capital flows (debt vs. equity based, public vs. private, and bank- vs. non-bank related) and more than two decades of quarterly data, they find that the global component of capital flows (which are in turn driven by both real and financial variables in their study) typically explains less than a quarter of the variation in actual cross-border capital flows, regardless of the type of capital flow or the measure used to define the global cycle, thus suggesting that domestic factors---including monetary policy, capital regulations, and others---remain key to financial and macroeconomic outcomes at a national level.

This paper revisits and expands this evidence by focusing on the sensitivity of national banking sectors to global factors in terms of a classic balance sheet ratio---namely the ratio of non-core to core liabilities---comparing such sensitivity in economies with flexible vs. fixed exchange rate regimes. As defined in a series of papers (see in particular \cite{shin2011procyclicality,shin2012global,hahm2013noncore,BrunoShin2015}), non-core bank funding consists of funding liabilities other than deposits, including repos, debt securities and foreign borrowing. For the consolidated banking system (once interbank domestic borrowing is netted out), this literature has also established  that foreign borrowing is a key component of such non-core liabilities, with cross-border borrowing (as well as lending) being driven by the activity of global banks which are in turn affected by cross-border financial conditions. Further, \cite{BrunoShin2015} have shown that cross-border borrowing has an important bearing on the leverage of national banking systems and on exchange rate movements.

We zoom in on the determinants of such a foreign liability component of banks’ non-core funding relative to core funding (which we henceforth generally call the non-core ratio), seeking to explain its dynamics as a function of domestic and global factors---both of real and financial natures. We further examine the sensitivity of the non-core to core ratios across countries, focusing on the long-standing Mundellian-inspired debate as to the extent to which exchange rate flexibility can help insulate a country’s banking system from the vagaries of global factors. Because of data limitations as well as to single out the role of exchange rate flexibility in countries with a fully open capital or financial account, our quarterly data sample spans only  advanced countries during the period 2004-2022. In doing so, another contribution of our empirical exercise lies in the proper identification of the global factors affecting national banking systems and relating such factors to specific financial and real variables at a global level, such as the VIX, world growth and regulation, the US short-term interest rate, the US real exchange rate, and oil prices. Such an identification of one or more global factors affecting banks is obtained  by combining the common correlated effect (CCE) estimator pioneered by \cite{pesaran2006estimation} with state-of-the-art approaches that permit identification of the relevant number of common factors and their relationship with the well-known observables mentioned above. 

As we discuss below, two important advantages of extracting principal components from residuals obtained by the CCE estimator are to allow (i) for the presence of unobserved or hard-to-measure global factors as well as (ii) for some cross-country heterogeneity in bank responses, which one might reasonably expect and which impart biases in the standard fixed effects estimator. We then establish how the choice of this approach influences our results relative to more standard panel fixed effects regressions in earlier studies. The end-goal is to establish the extent to which the estimated insulation role granted by  exchange rate flexibility is sensitive to econometric estimation (or model) uncertainty.

Our main findings are as follows. First, fluctuations in the ratio of non-core to core funding are highly persistent and overwhelmingly determined by global factors. Second, the combination of high cross-country dependence of non-core to core ratios and considerable heterogeneity in which global factors affect the banking system in each country, implies that the standard two-way fixed effects estimator (including country and time fixed effects) does not eliminate the cross-sectional dependence of the residuals, therefore producing likely inconsistent estimates of the response of bank ratios to the various country-specific and global disturbances. This indicates that the estimation methodology that we propose---combining the \citet{pesaran2006estimation} CCE estimator to extract the relevant global factors (principal components) and then using the mean group estimator to tease out any differences between countries on a fixed vs. flexible exchange rate regime---should produce superior results. This estimation methodology then delivers our central result, namely that countries on a fixed exchange rate regime are more susceptible to have global factors affecting their bank funding ratios, or in other words exchange rate flexibility helps insulate the non-core to core ratio from such global factors. We also find that countries with tighter macroprudential policy stances are less affected by the global financial cycle, and this insulation is most effective for countries with a fixed exchange rate or in a monetary union. We finally take a closer look at the mechanisms through which the principal components affect banks' non-core ratios. To this end, we first relate the principal components to the set of observed common factors, and find that most of the variation in banks' non-core ratios can be explained by changes in the world macroprudential policy stance, the US short-term interest rate and financial crisis effects. In contrast, the VIX, oil prices, and the US real exchange rate generally have a smaller impact on banks’ non-core ratios. Then, using BIS Locational Banking Statistic data, we break down total foreign borrowing by banks into (i) domestic- vs. foreign-currency borrowing, as well as (ii) liabilities provided by foreign banks vs. foreign non-bank financials vs. other foreign sectors. In these specifications, we find that our results are driven mostly by banks' interbank borrowing---a funding source that is known to be more flighty. The currency denomination does not seem to matter a lot.

The rest of this paper is structured as follows. In Section \ref{data}, we describe our data set and present its summary statistics. The empirical approach is laid out in Section \ref{approach}. Section \ref{results} presents the main results, while Section \ref{dissecting} studies the mechanisms through which the global financial cycle affects non-core ratios. Section \ref{conclusion} concludes the paper with a summary and brief discussion of our findings.

\section{Data}
\label{data}

Our empirical analysis is based on a data set that combines two critical banking sector balance sheet variables from the International Financial Statistics at quarterly frequency over the period 2004:Q1-2022:Q1 with different domestic macroeconomic controls and global variables. Our sample includes the 31 advanced economies listed in Table \ref{countrylist},\footnote{In most specifications, we drop Iceland from our sample as it is an outlier for most variables. Its inclusion reduces the significance of our estimates, as we document in the robustness checks of Section \ref{robsec}.} producing a balanced panel spanning more than 2,000 country-quarter observations. In this section, we give an overview about the data, with Appendix \ref{data_details} providing all of the corresponding data details and sources.

\subsection{Outcome Variables}

From the Other Depository Corporations Survey embedded in the IMF's International Financial Statistics (IFS), we obtain end-of-the-quarter data on two critical banking indicators---the non-core funding ratio for our baseline analysis as well as the loan-to-deposit ratio for robustness. The choice of these indicators follows \citet{jorda2017bank}, who identify them as important financial crisis predictors. \citet{shin2011procyclicality} and \citet{hahm2013noncore} put special emphasis on non-core funding ratios by showing that they can destabilize banking systems, especially when they mostly consist of foreign liabilities. More specifically, the latter two studies show that non-core liabilities tend to be a more elastic funding source than bank capital and deposits, thereby being closely correlated with faster credit growth relative to trend. So, one should expect to see in the data a significant  positive correlation between the ratio of non-core to core liabilities and the loan-to-deposit ratio. As noted in the introduction, \cite{hahm2013noncore} also show that banks’ foreign borrowing is a key component of non-core liabilities and especially elastic to credit growth.

Consistent with this evidence, we measure non-core funding ratios as the share of foreign liabilities over total deposits (IFS line 26c divided by IFS lines 24 and 25). To show how such a non-core ratio relates to other key balance sheet indicators highlighted in previous work, such as the ratio of bank loans-to-deposits (see \citealp{EdwardsVegh1997,CataoRodriguez2000,obstfeld2005trilemma,obstfeld2019tie}), we define loan-to-deposit ratio as IFS line 22d divided by IFS lines 24 and 25. As Table \ref{datatable} shows, the average non-core ratio in our sample is equal to 78\% and loan-to-deposit ratios have a mean of 122\%. The variations are substantial with standard deviations of about 94\% and 58\%, respectively. %The correlation between all three indicators is statistically significant at the 1\% level and equal to 29\%. 

\begin{figure}[htp!]
\centering%
	\begin{minipage}[b]{1\textwidth}%
		\begin{center}
			\begin{adjustbox}{max width=\textwidth}
				\begin{tabular}{c}
	\includegraphics[width=.75\textwidth]{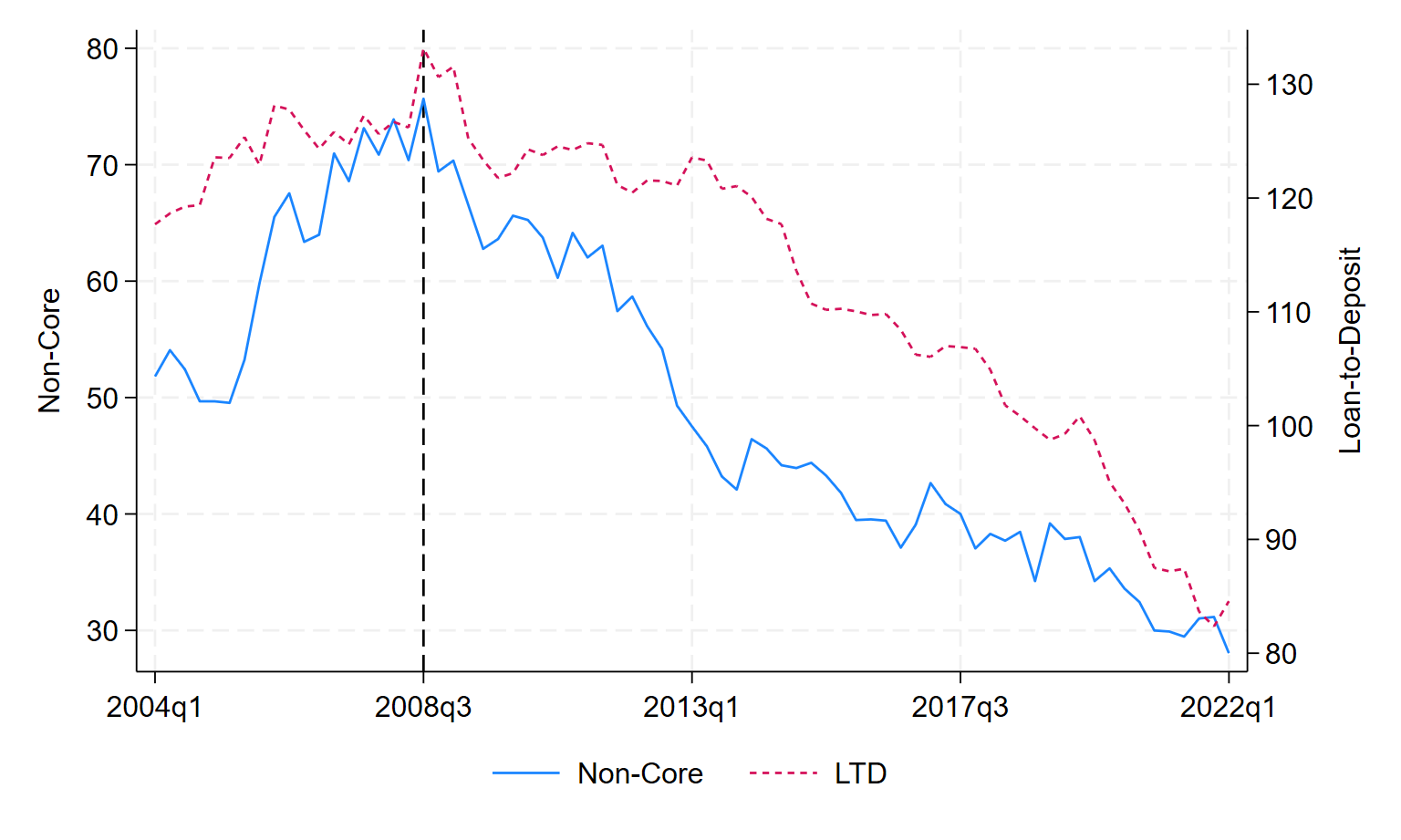} 
									\end{tabular}
			\end{adjustbox}
		\end{center}
  	\footnotesize{{\scshape Note}. This figure plots the median non-core ratio and loan-to-deposit ratio across all of the banking systems in our sample. The dashed vertical line marks the bankruptcy of Lehman Brothers in 2008:Q3. Sources: IFS, \cite{ilzetzki2019exchange}.}
	\end{minipage}
    \caption{\scshape Our Banking Indicators over Time}	\label{aggregatedynamics_all}
\end{figure}

%old Figure \ref{aggregatedynamics_all} depicts the time series dynamics of these three banking sector ratios. It becomes obvious that they all exhibit a distinct co-movement during our sample period. This strong correlation is not surprising given the evidence in \citet{hahm2013noncore}, who argue that capital and retail deposits are sticky and hence any increases in bank loans go along with greater interbank or foreign borrowing. This means that our measure of non-core ratios (foreign borrowing over deposits) should correlate with leverage (the ratio of bank loans to capital) and bank lending (loans over deposits).

\begin{figure}[htp!]
\centering%
	\begin{minipage}[b]{1\textwidth}%
		\begin{center}
			\begin{adjustbox}{max width=\textwidth}
				\begin{tabular}{c}
	\includegraphics[width=.8\textwidth]{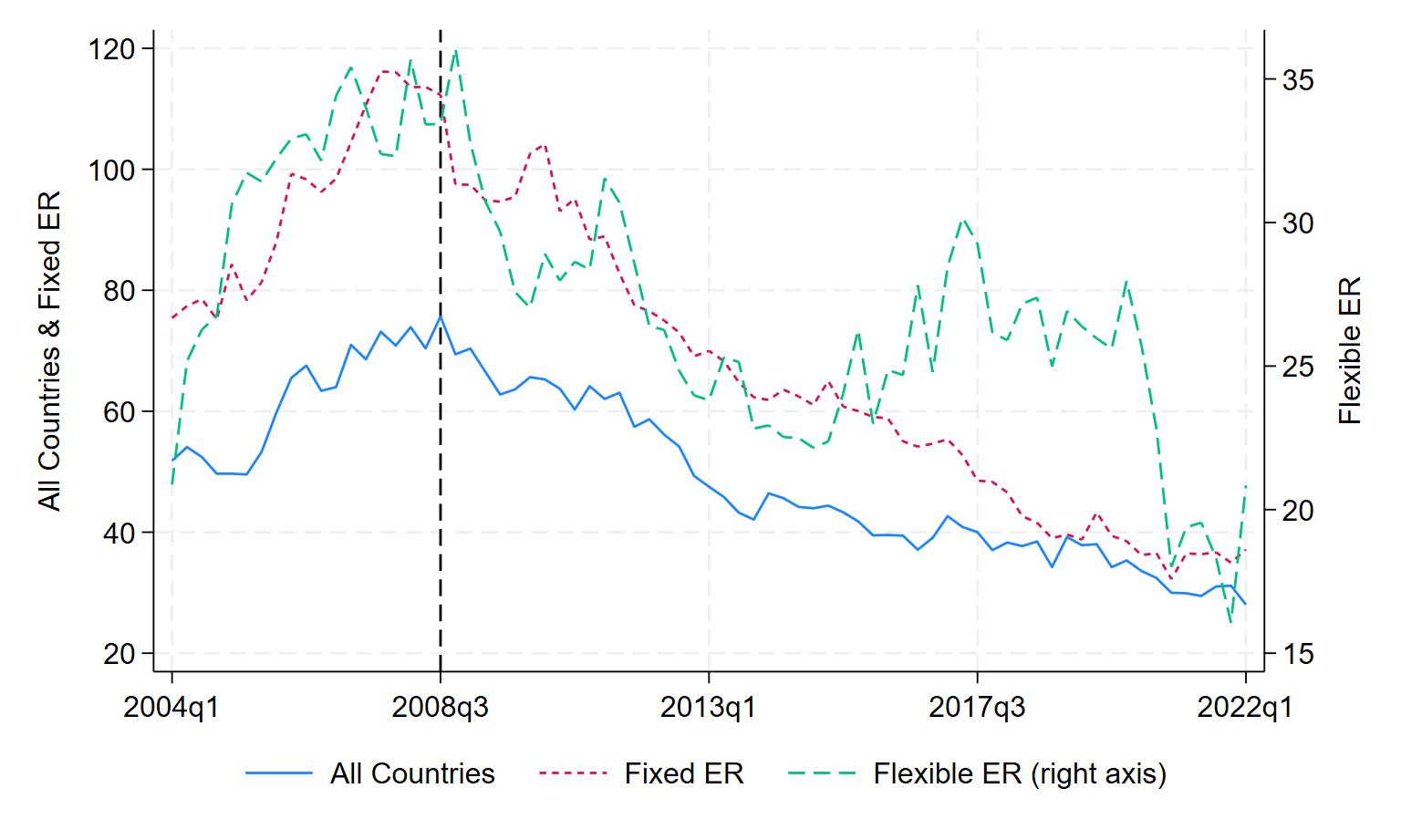} 
									\end{tabular}
			\end{adjustbox}
		\end{center}
  	\footnotesize{{\scshape Note}. This figure plots the median non-core funding ratio across all of the banking systems in our sample, as well as separately for those located in countries with a fixed vs. flexible exchange rate (evaluated with a one-quarter lag). The vertical dashed line marks the bankruptcy of Lehman Brothers in 2008:Q3. Sources: IFS, \cite{ilzetzki2019exchange}.}
	\end{minipage}
    \caption{\scshape The Non-Core Ratio over Time}	\label{aggregatedynamics}
\end{figure}

Figure \ref{aggregatedynamics_all} shows that advanced economies’ non-core funding ratios are positively correlated (\(\rho = 0.92\)) with swings in in banks’ loan-to-deposit ratios, revealing the important role of foreign borrowing in expanding lending beyond banks’ core funding sources (i.e., deposits). Figure \ref{aggregatedynamics} in turn shows that the non-core ratio is in general higher for countries on a fixed exchange rate.\footnote{Notice that the floaters are in the mean on the left scale that is steadily two thirds lower than its span.} Besides this level difference, it also shows conspicuous long-term trend shifts and considerable co-movement between the two country groups. In particular, non-core ratios increased significantly before the Global Financial Crisis, reaching a peak in 2008 where the global non-core ratio rose to 70\%, to then decrease to a level of 30\% in 2022. Only between 2015-20, there was some short-term divergence in the trends of fixers and floaters. %From that point on, some divergence between fixers and floaters is more apparent, with non-core ratios in countries with more flexible exchange rates recovering from the global financial crisis downfall.

In the second part of the paper, we examine in more detail the mechanisms through which banks' non-core funding ratios change. To this end, we employ data on banks' cross-border liabilities from the BIS Locational Banking Statistics, which we also scale by total deposits. The data have poorer country coverage than our IFS data, but have the advantage of allowing for a breakdown into domestic- vs. foreign-currency liabilities, as well as into foreign liabilities provided by banks, nonbank financials, and all other sectors. The correlation between our benchmark non-core ratio and the overall BIS cross-border liability variable is 88\%. Looking further into the different breakdowns of the BIS variable, we find that the correlation coefficient between the non-core ratio and the foreign-currency variable stands at 80\%, whereas it is 82\% for the domestic-currency variable, indicating that the
currency denomination of foreign liabilities does not seem to matter a lot in our sample. In addition, the non-core ratio correlates most tightly with cross-border liabilities vis-a-vis banks (89\%), followed by liabilities vis-a-vis nonbank financials (80\%) and all other sectors (18\%). These correlations suggest that our non-core ratio captures to large extents banks' interbank liabilities, an aspect that we will get back to in Section \ref{bis}.

 \begin{table}[tbh!]
 		\centering
 		\begin{adjustbox}{max height=.8\hsize, max width=.75\textwidth}
\begin{tabular}{l ccc cc}\hline\hline
 & Unit & Observations & $25th$ & Mean & $75th$ \\
 \hline  
 \multicolumn{6}{l}{Bank Variables}\\ \hline %
Non-Core Ratio & \% & 2263 & 22.09 & 78.00 & 94.75 \\
Loans-to-Deposit Ratio & \% & 2263 & 87.38 & 122.03 & 142.32 \\
Capital-to-Assest Ratio & \% & 2263 & 5.39 & 7.22 & 8.30 \\
BIS Cross-border Liabilities & \% & 1656 & 29.29 & 84.09 & 110.62 \\
BIS Foreign-currency, cross-border Liabilities & \% & 1651 & 10.75 & 39.78 & 58.72 \\
BIS Domestic-currency, cross-border Liabilities & \% & 1651 & 15.00 & 44.29 & 61.69 \\
BIS Cross-border Liabilities to Banks & \% & 1579 & 18.39 & 53.98 & 69.71 \\
BIS Cross-border Liabilities to Nonbanks & \% & 1656 & 6.14 & 20.99 & 28.57 \\
BIS Cross-border Liabilities to Other Sectors & \% & 1579 & 0.00 & 9.06 & 11.93 \\
\hline \multicolumn{6}{l}{Domestic Variables}\\ \hline 
GDP Growth & \% & 2263 & -2.69 & 0.57 & 4.33 \\
Real Interest Rate & pp & 2263 & -2.28 & -0.25 & 1.80 \\
Real Exchange Rate & - & 2263 & 93.48 & 98.45 & 101.55 \\
Capital Regulation & - & 2263 & 0.00 & 0.51 & 0.67 \\
Liquidity Regulation & - & 2263 & 0.00 & 0.35 & 0.33 \\
Money / GDP & \% & 2263 & 74.03 & 94.82 & 104.02 \\
\hline \multicolumn{6}{l}{Global Variables}\\ \hline 
VIX & - & 2263 & 2.62 & 2.87 & 3.05 \\
Oil Price & USD & 2263 & 49.14 & 67.79 & 88.01 \\
Shadow Rate & \% & 2263 & -1.20 & 0.69 & 2.03 \\
Non-Bank Share & \% & 2263 & 46.52 & 48.25 & 49.84 \\
World GDP Growth & \% & 2263 & -3.01 & 0.50 & 2.57 \\
World Macroprud. & - & 2263 & 0.22 & 1.74 & 3.55 \\
World Inflation & \% & 2263 & 0.20 & 0.45 & 0.82 \\
Real USD Exchange Rate & - & 2263 & 83.79 & 90.94 & 96.54 \\ \hline
\end{tabular}
	\end{adjustbox}
 \caption{\scshape Summary Statistics}
 	\label{datatable}
 \justify  \footnotesize{{\scshape Note}. The table reports selected summary statistics. See Table \ref{datasources} for data definitions and sources. }
 \end{table}

\subsection{Explanatory Variables}

Our empirical analysis rests on a sizeable set of variables that both theory and previous empirical work indicate to be potentially important to explain lending and funding decisions by banks. It is conceptually appealing for the task at hand to group such variables into two sub-sets, country-specific (domestic) and global ones. 

\subsubsection{Domestic Variables}

The first comprises variables that are more directly controlled by national economic policy and/or largely shaped by country-specific institutions and structural characteristics. These include the domestic short-term real interest rate, the real exchange rate, the ratio of broad money to GDP, bank and macro-prudential regulations, as well as real GDP growth. 

Banking models---both those embedded in general equilibrium (\citealp{EdwardsVegh1997,CataoRodriguez2000,CespedesChangGarciaCicco2011}), as well as partial equilibrium ones (e.g., \citealp{BrunoShin2015})---show that increases in the (real) domestic interest rate and appreciations of the real exchange rate tend to favor foreign funding relative domestic-deposit funding, whereas liquidity and minimum capital requirements tend to depress lending and hence external borrowing by banks, all else constant.\footnote{Considering a variety of regulatory instruments that include liquidity, loan-to-value and minimum capital rules, \cite{alam2019digging} document sizeable effects on bank lending and private consumption growth in a cross-country panel dataset.} To take into account the relevance of regulation, we control for (i) the country-level capital-to-asset ratio of banks and (ii) the tightness of macroprudential liquidity and capital regulation, based on the IMF's integrated Macroprudential Policy (iMaPP) database, originally constructed by \citet{alam2019digging}. Specifically, the database reports for each country in the sample whether it tightens or loosens the degree of capital or liquidity regulation, respectively. As macroprudential regulation is likely to have a longer-lasting effect on bank balance sheets, we cumulate the two variables for each country.

Conversely, one might expect faster-growing economies to be more reliant on less inelastic funding sources like cross-border borrowing, thereby potentially raising non-core bank funding---this positive effect is likely reinforced by evidence that faster GDP growth lowers the share of non-performing loans in banks’ portfolio, thereby raising their non-deposit funding capacity (see \cite{AriChenRatnovski20219}). Measures of domestic financial deepening can also be expected to play a role due to the reasons mentioned in \cite{BrunoShin2015} on how higher ratios of broad money to GDP correlate positively with non-core funding by reflecting more extensive private sector tapping of foreign finance. 

In order to study the extent to which the Mundellian trilemma---which postulates that it is impossible for a country to simultaneously have a stable exchange rate, full capital mobility, and an independent monetary policy (\citealp{mundell1963capital}; \citealp{obstfeld2005trilemma})---affects the relation between the global financial cycle and bank balance sheets, we merge an exchange rate dummy equal to one for fixed exchange rate regimes (i.e., strictly fixed pegs), following the classification of \cite{ilzetzki2019exchange}, to our dataset. Our sample is roughly balanced between countries with fixed and flexible exchange rates, with 53\% of the countries having a fixed exchange rate, which includes membership of a monetary union. As the average Chinn-Ito index (\citealp{chinn2006matters}) in our sample stands at 0.95, our data set essentially consists of countries with open financial accounts only, and we hence do not explicitly control for this variable.\footnote{Including this variable leaves our main results largely unchanged.} This is chiefly an advantage of using a sample including only advanced economies, since, when examining the predictions of the Mundellian trilemma, one can focus squarely on the role of the exchange rate regime. We winsorize all country-specific, independent variables at the 1\% and 99\% levels to limit the potential impact of outliers.

\subsubsection{Global Variables}

The other sub-set of explanatory variables comprises global factors. Those include fluctuations in risk aversion in global capital markets, world oil prices, the world non-bank share, weighted averages of inflation, GDP growth and macroprudential tightness of all of the economies in our sample, as well as the US shadow interest rate of \citet{wu2016measuring},\footnote{Relative to the US Federal Funds rate, this shadow rate has the advantage of not being bound at zero, so that the effects of QE can still be taken into account.} and the dollar valuation relative to other currencies---the latter two reflecting the still hegemonic roles of the US financial system and of the US dollar in global finance and trade. \cite{BrunoShin2015} as well as \cite{CoeurdacierReyWinant2020}, \cite{diGiovanniKalemi2022} and \cite{obstfeld2019tie} underline the role of rising global risk aversion (as measured by the VIX) in dampening cross-border banking borrowing, hence potentially affecting the non-core to core ratio. In their survey of existing empirical work on the global financial cycle, \cite{MIRANDAAGRIPPINO20221} emphasise that tighter US monetary policy tends to depress global asset prices and discourage global financial intermediation, with oil prices and other real global factors (proxied by global GDP and trade growth) also playing a role in international fluctuations in global bank credit (and hence banks' funding decisions). Other recent studies have emphasized the role of US dollar valuations in global capital flows and financial intermediation broadly defined (\citealp{avdjiev2019dollar}; \citealp{jiang2020dollar}), so it seems sensible to add a trade-weighted index of the US dollar exchange rate as another potential global factor. We finally saturate the regressions with the worldwide average share of financial intermediation performed by nonbanks, assuming that a higher world non-bank share correlates with lower credit volumes and hence foreign borrowing of banks, as well as the GDP-weighted average of advanced economies' cumulative macroprudential tightness, as we expect a tight link between global macroprudential tightness and banks' funding choices. The latter variable is again based on the iMaPP Database.\footnote{Breaking down this overall world macroprudential stringency index into the capital and liquidity sub-components leaves our results unchanged. Attendant results are available upon request.} Table \ref{datatable} provides the descriptive statistics for these global variables. 

Overall, one might expect considerable correlations between the two groups of variables---country-specific and global ones---so it is important that the econometric approach employed to identify country vs. global factors allows for these correlations. In the next section, we explain how we do this. 

\section{Empirical Approach}
\label{approach}

In this section, we outline the econometric model. We will start with a generic model and then turn to the specification tailored to our specific testing procedures. Denoting the number of countries with \(N\) and the number of time periods with \(T\), an interactive fixed effects (IFE) model with unit-specific slopes and observed and unobserved heterogeneity is:
\begin{align}
%    y_{i,t} &=  \rho_i y_{i,t-1} + \boldsymbol{x}_{i,t} \boldsymbol{\beta}_i +\boldsymbol{g}_t  \boldsymbol{\delta}_i+ v_{i,t} \label{eq:PanelModel}\\
    y_{i,t} &=  \rho_i y_{i,t-1} + \boldsymbol{x}_{i,t} \boldsymbol{\beta}_i + v_{i,t} \label{eq:PanelModel}\\
    v_{i,t}  &= \boldsymbol{f}_t \boldsymbol{\gamma}_i  + \epsilon_{i,t},     
\end{align}
where \(y_{i,t}\) is the dependent variable and \(y_{i,t-1}\) is its one-quarter lag. \(\boldsymbol{x}_{i,t}\) is a \(1\times K\) vector of  strictly exogenous explanatory variables.  %\(\boldsymbol{g}_t\) are the \(M_{ocf}\) observed common factors, such as global oil prices, inflation and the VIX, and 
\(\boldsymbol{f}_t\) are the \(M_{ucf}\) unobserved common factors, modelled as a part of the overall unobserved multifactor error \(v_{i,t}\) with the iid random noise term  \(\epsilon_{i,t}\).\footnote{To avoid strong endogeneity of the right-hand side variables, we use one-quarter lags of all observable variables.} Common factors affect all countries; however, due to the slope heterogeneity in the loadings %\(\boldsymbol{\delta}_i\) and 
\(\boldsymbol{\gamma}_i\), the factors affect countries with a different strength.\footnote{The additive fixed effects model is nested within the interactive fixed effects model if we assume the first element in \(\boldsymbol{f}_t\) and for the two-way fixed effect model the second element of \(\boldsymbol{\gamma}_i\) to be one.} The vector of unobserved coefficients {\(\boldsymbol{\beta}_i=[\beta_{i,1},...,\beta_{i,K}]'\) }is allowed to vary across countries. 

The advantage of the IFE model is that, in comparison to the widely used additive or two-way fixed effects model, it allows for the presence of unobserved common shocks, which can be correlated with the explanatory variables. For example, assume that the country-specific vector of observed exogenous variables  \(\mathbf{x}_{i,t}\) is driven by %the observed factors \(\boldsymbol{g}_t\) and 
unobserved common factors \(\boldsymbol{f}_t\):
\begin{align}
    %\mathbf{x}_{i,t}= \boldsymbol{g}_t \boldsymbol{\delta}_i^x+\boldsymbol{f}_t \boldsymbol{\gamma}_i^x+\xi_{i,t}, \label{eq:EqX}
    \mathbf{x}_{i,t}= \boldsymbol{f}_t \boldsymbol{\gamma}_i^x+\xi_{i,t}, \label{eq:EqX}
\end{align}
then ignoring the %observed and 
unobserved common factors will cause strong cross-sectional dependence, which can lead to biased and inconsistent estimates of \(\mathbf{\beta}_i\). 

To establish the presence of either factors, their number can be estimated \citep{Bai2002,AhnHorenstein2013}. Alternatively, a test for strong cross-sectional dependence, the so called CD test \citep{PesaranCD2015}, can be performed.\footnote{For an intuitive definition of weak and strong cross-section dependence, see \cite{Ditzen2018xtdcce2}.} The null hypothesis of the CD test is weak cross-sectional dependence while the alternative is strong dependence. The test statistic is a weighted average of the estimated pairwise correlations.

If the unobserved common factors are correlated with the explanatory variables and omitted from the model, an omitted variable bias occurs. The most popular estimation methods in the literature to estimate models in the presence of strong cross-sectional dependence originating from unobserved common factors are the  \textit{Common Correlated Effects} (CCE) estimator \citep{pesaran2006estimation} or \textit{Principal Components} (PC) estimator \citep{Bai2009,Moon2015}. The CCE estimator proved to be very flexible under different assumptions on the errors \citep{ChudikPesaranTosetti2011}, error structure \citep{KapetaniosPesaranYamagata2011} or exogeneity \citep{ChudikPesaranJoE2015}. Further, knowledge about the exact number of unobserved common factors is not required for the CCE estimator, which is an advantage over the PC estimator. The unobserved common factors are approximated by time-specific averages across all units, the so-called cross-sectional averages. The estimator is consistent if the number of unobserved common factors is smaller than the number of cross-section averages and if \(N,T \rightarrow \infty\) with \(T/N\rightarrow 0\). The estimated equation using the CCE estimator becomes:
\begin{align}
   % y_{i,t} &= \rho_i y_{i,t-1} + \mathbf{x}_{i,t} \boldsymbol{\beta}_i + \boldsymbol{g}_t\boldsymbol{\delta}_i+\sum_{l=0}^p \bar{\boldsymbol{R}}_{t-l}\boldsymbol{\gamma}_{i,l}+e_{i,t} \label{eq:mainCCE}\\
    y_{i,t} &= \rho_i y_{i,t-1} + \mathbf{x}_{i,t} \boldsymbol{\beta}_i + \sum_{l=0}^p \bar{\boldsymbol{R}}_{t-l}\boldsymbol{\gamma}_{i,l}+e_{i,t} \label{eq:mainCCE}\\
    \bar{\boldsymbol{R}}_t &= \frac{1}{N}\sum_{i=1}^N \boldsymbol{R}_{i,t} = \frac{1}{N}\sum_{i=1}^N  \left[y_{i,t},\mathbf{x}_{i,t} \right],
\end{align}
where \(\bar{\boldsymbol{R}}_t\) are the cross-section averages of the dependent and independent variables. Since Equation \eqref{eq:mainCCE} includes lags of the dependent variable on the right-hand side, \(p\) lags of the cross-section averages are added to approximate the unobserved common factors \citep{PesaranChudik2019a}. \(e_{i,t}\) is the iid regression residual.

The CCE estimator can also be combined with the mean group or pooled estimator. The mean group estimator \citep{PesaranShinSmith1999,PesaranChudik2019a} allows for slope heterogeneity across units and averages the individual coefficients.  The pooled CCE estimator assumes homogeneous coefficients, that is the same causal effect for all cross-sectional units, both in the short and long run in the case of dynamic models.\footnote{Note, however, that the widely-used Pooled Mean Group (PMG) Estimator \citep{PesaranShinSmith1999} assumes long-run homogeneity only in the long-run coefficients, rather than both in the short and long run. In all regression specifications, the pooled CCE estimator failed to remove cross-country residual correlation and was therefore inferior to the MG estimator.}

Moving on to the specifics of our model, denoting the noncore-to-core ratio as the dependent variable, we distinguish the observed variables into country-specific variables and \(M_{ocf}\) observed common factors, \(\boldsymbol{g}_t\), such as global oil prices, inflation and the VIX: 
\begin{align}
    y_{i,t} &= \rho_i y_{i,t-1} +\mathbf{x}_{i,t}\boldsymbol{\beta}_i +  \boldsymbol{g}_t\boldsymbol{\theta}_i +  v_{i,t} \label{eq:emp1}\\ 
    v_{i,t} &= \boldsymbol{\gamma}_i\boldsymbol{f}_t + \epsilon_{i,t}\label{eq:emp2}.
\end{align}

We further assume that the observed country-specific variables are driven by both types of common factors, \(\boldsymbol{f}_t\) and \(\boldsymbol{g}_t\), and the observed common factors, \(\boldsymbol{g}_t\), are exposed to unobserved shocks as well:
\begin{align}
    \boldsymbol{x}_{i,t} &= \boldsymbol{\gamma}_i^x\boldsymbol{f}_t+\boldsymbol{g}_t\boldsymbol{\theta}_i^x + \boldsymbol{\chi}_{i,t} \label{eq:emp3}\\ 
     \boldsymbol{g}_{t} &= \boldsymbol{\xi}\boldsymbol{f}_t+ \boldsymbol{\omega}_{t}\label{eq:emp4}.
\end{align}

The term \(\boldsymbol{\chi}_{i,t}\) is driven neither by observed nor unobserved common factors. \(\boldsymbol{\omega}_{t}\) is the variation of the observed common factors that is independent of the unobserved common factors. Our main aim is the identification of \(\boldsymbol{\chi}_{i,t}\) and \(\boldsymbol{\omega}_{t}\). However, only \(\boldsymbol{x}_{i,t}\) and \(\boldsymbol{g}_{t}\) are observed. %Therefore, ignoring the unobserved common factors will lead to an omitted variable bias. 

Substituting \eqref{eq:emp2} and \eqref{eq:emp3} and \eqref{eq:emp4} into \eqref{eq:emp1} and collecting terms leads to:

\begin{align}
y_{i,t} =& \rho_i y_{i,t-1} +\boldsymbol{\beta}_i\boldsymbol{\chi}_{i,t} + \boldsymbol{\omega}_{t}\left(\boldsymbol{\beta}_i\boldsymbol{\theta}_i^x + \boldsymbol{\theta}_i\right) \label{eq:FinalModel}\\
&+ \boldsymbol{f}_t\left[\boldsymbol{\beta}_i\left(\boldsymbol{\gamma}_i^x+\boldsymbol{\theta}_i^x\boldsymbol{\xi}\right)+\boldsymbol{\gamma}_i+\boldsymbol{\theta}_i\boldsymbol{\xi}\right]+\epsilon_{i,t}. \nonumber
\end{align}

The first two terms represent the effect of the country-specific variables and of the observed common factors on the dependent variable. The third term is the effect of the unobserved common factors. Although the observed common factors are known and can be directly used, the unobserved common factors will be approximated by cross-section averages. We further define a composite, \(u_{i,t}\), which contains the observed and unobserved common factors:

\begin{align}
u_{i,t} &= \boldsymbol{\omega}_{t}\left(\boldsymbol{\beta}_i\boldsymbol{\theta}_i^x + \boldsymbol{\theta}_i\right)+ \boldsymbol{f}_t\left[\boldsymbol{\beta}_i\left(\boldsymbol{\gamma}_i^x+\boldsymbol{\theta}_i^x\boldsymbol{\xi}\right)+\boldsymbol{\gamma}_i+\boldsymbol{\theta}_i\boldsymbol{\xi}\right] +\epsilon_{i,t} \label{eq:u}
\intertext{and Equation \eqref{eq:FinalModel} then becomes:}
y_{i,t} &= \rho_i y_{i,t-1} +\boldsymbol{\beta}_i\boldsymbol{\chi}_{i,t} + u_{i,t}.
\end{align}

Our estimation approach consists of three steps. The first step is a CCE-MG estimation of \eqref{eq:FinalModel} where the observed common factors are directly added and the cross-section averages span the remaining part of the factor space:
\begin{align}
  y_{i,t} &= \rho_i y_{i,t-1} + \mathbf{x}_{i,t} \boldsymbol{\beta}_i + \boldsymbol{g}_t\boldsymbol{\theta}_i+\sum_{l=0}^p \bar{\boldsymbol{R}}_{t-l}\boldsymbol{\gamma}_{i,l}+e_{i,t}. \label{eq:CCEModel}  
\end{align}

An advantage of the CCE approach is that it incorporates the observed common factors in the same way as the observed country-specific variables. Therefore, as long as the number of cross-section averages exceeds the number of unobserved common factors, the coefficients \(\boldsymbol{\beta}_i\) can be consistently estimated. The first step then approximates the unobserved common factors  \(\boldsymbol{f}_t\) with the cross-section averages, while the observed factors \(\boldsymbol{g}_t\) are directly used. Implicitly, we are using the variation in \(\mathbf{x}_{i,t}\), which is independent of the observed and unobserved common factors, \(\boldsymbol{\chi}_{i,t}\), to obtain an estimate of \(\boldsymbol{\beta}_i\). We then estimate the composite term \(u_{i,t}\) from Equation \eqref{eq:u}:
\begin{align}
\hat{u}_{i,t} &=  y_{i,t} - \hat{\rho}_i y_{i,t-1} - \mathbf{x}_{i,t} \boldsymbol{\hat{\beta}}_i. \label{eq:uhat}
\end{align}
Yet, our aim is the identification of the interaction of the observed common factors and the country-specific variation \(\boldsymbol{\chi}_{i,t}\), like the exchange rate regime and other policy or regulatory parameters. To identify \(\hat{u}_{i,t}\) correctly, we must ensure that no common factors or cross-section dependence is left in \(\epsilon_{i,t}\). 

In the second step, we therefore extract the common components of the observed common factors, of the unobserved ones, and of the remaining residual variation from \(\hat{u}_{i,t}\) by using principal components. The number of components is determined by the criteria of \cite{AhnHorenstein2013}. In the third step, we then replace the observed and unobserved common factors with the principal components from the previous step and estimate the following model:
\begin{align}
   y_{i,t} = \rho_i y_{i,t-1} + \mathbf{x}_{i,t} \boldsymbol{\beta}_i +\boldsymbol{PC}_t \boldsymbol{\kappa}_i    + \zeta_{i,t},
\end{align}
where \(\boldsymbol{PC}_t\) are the first \(p_{pc}\) principal components of the observed and unobserved common factors and of the estimated residuals and \( \zeta_{i,t}\) is iid and uncorrelated across units.\footnote{Our model is related to the regularized CCE estimator in \cite{Juodis2022}. The idea is to reduce the dimension of the cross-section averages to the dimension of the estimated number of common factors.}

\section{Results}
\label{results}

\subsection{Benchmark Results}

\begin{table}%
%\centering
\tiny
\resizebox{\textwidth}{!}{\begin{tabular}{l cc cc cc c }\hline\hline
                              &         (1)   &         (2)   &         (3)   &         (4)   &         (5)   &         (6)   &         (7)   \\
                              &        TWFE   &          FE   &      TWFE-I   &          MG   &      MG-CCE   &          MG   &          MG   \\
\hline
Lag Dep. Var.              &       0.986   &       0.986   &       0.980   &       0.715   &       0.541   &       0.659   &       0.659   \\
                              &     (0.016)***&     (0.017)***&     (0.018)***&     (0.030)***&     (0.035)***&     (0.031)***&     (0.031)***\\
GDP Growth                     &       0.015   &       0.017   &       0.019   &      -0.049   &       0.019   &      -0.013   &      -0.013   \\
                              &     (0.038)   &     (0.039)   &     (0.038)   &     (0.159)   &     (0.172)   &     (0.091)   &     (0.091)   \\
Real Interest Rate                    &      -0.039   &      -0.014   &      -0.020   &      -0.182   &      -0.313   &      -0.065   &      -0.065   \\
                              &     (0.056)   &     (0.055)   &     (0.062)   &     (0.083)** &     (0.306)   &     (0.072)   &     (0.072)   \\
REER                      &       0.098   &       0.106   &       0.067   &       0.704   &       0.659   &       0.424   &       0.424   \\
                              &     (0.025)***&     (0.025)***&     (0.026)** &     (0.284)** &     (0.256)** &     (0.186)** &     (0.186)** \\
Capital-to-Asset             &      -0.217   &      -0.216   &      -0.264   &      -0.135   &      -0.303   &       0.807   &       0.807   \\
                              &     (0.156)   &     (0.154)   &     (0.160)*  &     (0.590)   &     (0.940)   &     (0.482)*  &     (0.482)*  \\
Capital Reg.                &      -0.292   &      -0.203   &      -0.495   &      -0.652   &      -3.274   &      -4.051   &      -4.051   \\
                              &     (0.357)   &     (0.358)   &     (0.375)   &     (1.765)   &     (2.321)   &     (2.908)   &     (2.908)   \\
Liquidity Reg.                 &       0.634   &       0.489   &       0.524   &       0.348   &       1.541   &      -0.305   &      -0.305   \\
                              &     (0.444)   &     (0.438)   &     (0.465)   &     (2.281)   &     (4.564)   &     (2.977)   &     (2.977)   \\
Money / GDP                 &       0.046   &       0.012   &       0.039   &      -0.021   &       0.253   &       0.176   &       0.176   \\
                              &     (0.017)***&     (0.018)   &     (0.015)***&     (0.051)   &     (0.187)   &     (0.064)***&     (0.064)***\\
\multicolumn{6}{l}{Observed Common Factors} \\ \ VIX                       &               &       0.342   &               &      -0.030   &      -0.945   &               &               \\
                              &               &     (0.787)   &               &     (0.859)   &     (1.285)   &               &               \\
\ Oil Price                  &               &      -0.001   &               &      -0.007   &       0.001   &               &               \\
                              &               &     (0.020)   &               &     (0.028)   &     (0.048)   &               &               \\
\ Shadow Rate                &               &       0.301   &               &       0.400   &      -0.154   &               &               \\
                              &               &     (0.113)***&               &     (0.366)   &     (0.419)   &               &               \\
\ Non-Bank Share              &               &       0.326   &               &      -0.383   &       0.584   &               &               \\
                              &               &     (0.284)   &               &     (0.306)   &     (0.492)   &               &               \\
\ World GDP Growth               &               &       0.128   &               &       0.098   &       0.060   &               &               \\
                              &               &     (0.084)   &               &     (0.141)   &     (0.312)   &               &               \\
\ World Macropru.               &               &      -0.468   &               &      -0.675   &       0.188   &               &               \\
                              &               &     (0.253)*  &               &     (0.590)   &     (1.779)   &               &               \\
\ World Inflation            &               &       0.451   &               &       0.360   &       0.579   &               &               \\
                              &               &     (0.422)   &               &     (0.651)   &     (1.166)   &               &               \\
\ US REER                   &               &       0.000   &               &      -0.023   &       0.043   &               &               \\
                              &               &     (0.089)   &               &     (0.108)   &     (0.130)   &               &               \\
\multicolumn{6}{l}{Principal Components (PC, fixed = 0 if coefficients vary)}\\ \ PC1                        &               &               &               &               &               &       1.285   &       0.302   \\
                              &               &               &               &               &               &     (0.527)** &     (0.163)*  \\
\ PC2                        &               &               &               &               &               &       0.709   &      -0.080   \\
                              &               &               &               &               &               &     (0.400)*  &     (0.156)   \\
\ PC3                        &               &               &               &               &               &       1.194   &       0.373   \\
                              &               &               &               &               &               &     (0.608)** &     (0.204)*  \\
\multicolumn{6}{l}{Principal Components (PC, fixed = 1) }\\ \ PC1                        &               &               &               &               &               &               &       0.984   \\
                              &               &               &               &               &               &               &     (0.521)*  \\
\ PC2                        &               &               &               &               &               &               &       0.789   \\
                              &               &               &               &               &               &               &     (0.363)** \\
\ PC3                        &               &               &               &               &               &               &       0.821   \\
                              &               &               &               &               &               &               &     (0.591)   \\
\hline  Obs                   &        2160   &        2160   &        2160   &        2160   &        2160   &        2160   &        2160   \\
Countries                     &          30   &          30   &          30   &          30   &          30   &          30   &          30   \\
Time Periods                  &          72   &          72   &          72   &          72   &          72   &          72   &          72   \\
\hline CD                     &      10.583   &      13.531   &       5.575   &      12.699   &       0.797   &      -0.134   &      -0.134   \\
\ p                           &       0.000   &       0.000   &       0.000   &       0.000   &       0.425   &       0.894   &       0.894   \\
\hline \(R^2\)                &       0.992   &       0.992   &       0.993   &       0.040   &       0.173   &       0.042   &       0.042   \\
\(R^2\) (MG)                  &               &               &               &       0.948   &       0.957   &       0.950   &       0.947   \\
\(R^2\) (within)              &       0.911   &       0.932   &       0.894   &       0.960   &       0.053   &       0.958   &       0.958   \\
\hline \(\hat{M}_{\hat{u}}\)&               &               &               &               &           2   &               &               \\
\(\hat{M}_{\hat{\epsilon}}\)&           2   &           2   &           2   &           0   &           0   &           0   &           0   \\\hline\hline

\end{tabular}}
\justify{This table displays the results for the years 2004-2022 with the non-core ratio as dependent variable. Standard errors in parenthesis. Significance levels: \(^* 10\%, ^{**}5\%, ^{***}1\%\). 
TWFE and TWFE-I include country and time fixed effects. TWFE-I additionally includes interactions between the fixed exchange rate dummy and time fixed effects. All explanatory variables except  Capital Regulation and  Liquidity Regulation are added as cross-section averages for MG-CCE. Principal Components (PC) are extracted from composite \(\hat{u}_{i,t}\) in column MG-CCE. \textit{CD} is the value of the CD test \citep{PesaranCD2015} with the null hypothesis of weak cross-section dependence. \(\hat{M}_{X}\) is the estimate of the number of common factors using the Growth Criterion from \cite{AhnHorenstein2013}. \(\hat{u}\) is the composite of observed and unobserved common factors from Equation \eqref{eq:uhat} and \(\hat{\epsilon}\) the residual. All variables are defined as in Table \ref{datasources}.}
 \caption{\scshape Benchmark Results: 2004-2022}
 \label{tab:MainResults}
\end{table}

\begin{table}%
%\centering
\tiny
\begin{tabular}{l cc cc  }\hline\hline
& (4) & (5) & (6) & (7) \\
            &          MG&      MG-CCE&          MG&          MG\\
\hline F: PC = 0&            &            &      11.367&       7.120\\
\ p         &            &            &       0.000&       0.000\\
\(\Delta\) (all)&      18.249&      14.406&      62.249&      53.578\\
\ p         &       0.000&       0.000&       0.000&       0.000\\
\(\Delta\) (PC)&            &            &      -7.006&     -10.232\\
\ p         &            &            &       0.000&       0.000\\
t: \(PC F_1 > PC F_0\)&            &            &            &       5.183\\
\ p         &            &            &            &       0.011\\
\hline Share PC (all)&            &       0.795&            &            \\
\ PC 1      &            &       0.585&            &            \\
\ PC 2      &            &       0.138&            &            \\
\ PC 3      &            &       0.073&            &            \\\hline\hline

\end{tabular}
\justify{Additional Results corresponding to columns (4), (5), (6) and (7) of Table \ref{tab:MainResults}.  \(F: PC = 0\) tests if the coefficients of the principal components are joint significantly different from zero. \(\Delta (all)\) test the null of slope homogeneity of all coefficients, \(\Delta (PC)\) test the null of slope homogeneity of the coefficients of the principal components under the assumption that all other coefficients are heterogeneous. \(PC F_1 > PC F_0\) test if the sum of the coefficients of the PCs for \(fixed = 1\) are larger than those for \(fixed=0\). Share PC is the share of the explained variance of the PC. All variables are defined as in Table \ref{datasources}.}
 \caption{\scshape Benchmark Results: 2004-2022}
 \label{tab:MainResultsAdd}
\end{table}

The existence of a global factor entails some commonality of movements across units (countries in our case) and failing to take that into proper account in the econometric estimation can bias the estimates, so it is crucial that we test each panel regression for the presence of strong cross-sectional dependence using the CD test before making any inference. Additionally, we want to ensure that  the entire factor structure due to the observed and unobserved common factors from the dependent variable is captured in the composite \(\hat{u}_{i,t}\). If the factor structure is fully enclosed in the composite, the remaining variation used to estimate the coefficients of the exogenous country-specific variables is independent of any factors. This further ensures an unbiased estimation. In the second step of our three-stage estimation procedure, we obtain principal components from the the composite \(\hat{u}_{i,t}\) to approximate the factor structure in the dependent variable. Therefore, missing any factors in the composite would invalidate the second and third step of our estimation procedure described in the previous section.

Turning to the data, we first estimate the number of common factors in the dependent variable, the non-core ratio, to be 3.\footnote{There are 7 factors in the observed common factors, see Table \ref{tab:NumComFac}. This implies that some of the observed factors have no or only minor relevance for the non-core ratio.} We will use this number as the number of common factors. To ensure the factor structure is fully captured, we compare this number to the number of factors in the composite, \(\hat{M}_{\hat{u}}\), and the residual, \(\hat{M}_{\hat{\epsilon}}\). %We will use this number as an indicator whether we correctly account for all factors in the composite.
 
The first column of Table \ref{tab:MainResults} presents a standard two-way fixed effects specification with homogeneous coefficients, country and year fixed effects, and no accounts of any observed or unobserved common factors.\footnote{We use Stata 18 and the community contributed packages \textit{reghdfe}  \citep{reghdfe}  for the TWFE and FE estimations and \textit{xtdcce2} and \textit{xtcd2} \citep{Ditzen2018xtdcce2,Ditzen2021} for the MG-CCE regressions and  the CD tests. The number of common factors is estimated using \texttt{xtnumfac} \citep{ReeseDitzen2023}.} The CD test rejects the null of weak cross-section dependence, i.e., it suggests that there are global factors that are not accounted for and/or those global factors are affecting countries in a distinct way.\footnote{\cite{JuodisReese2021} show that the CD test statistic diverges under certain conditions and propose a corrected test. Our results are qualitatively similar using the their test. } This is supported by two estimated common factors in the residuals (\(\hat{M}_{\hat{\epsilon}}\)). Moving rightwards along the table, we attempt to account for the factor structure in two ways. In column (2), we add the observed common factors and in column (3), we interact the time fixed effects with the fixed exchange rate indicator. Two of the observed common factors are now significant in column (2), but even in this case and in column (3) when interacting the time fixed effects with the exchange rate indicator, strong cross-section dependence remains in the residuals, indicating a potential bias of the regressions results. 

Moving to columns (4) and (5), we change our specification in two aspects. First, we allow for country-specific coefficients on both the country-specific variables and observed global factors using the mean group estimator (MG). The test for slope heterogeneity, presented in the columns labelled (4) and (5) in Table \ref{tab:MainResultsAdd} confirms the necessity to allow for country-specific slopes.\footnote{The tests for slope heterogeneity were performed with the \textit{xthst} package \citep{BersvendsenDitzen2021}.}  Turning to the results in column (4) of Table \ref{tab:MainResults}, while no factors remain in the residuals, the CD test still rejects the null of weak cross-section dependence. Further, none of the observed common factors are significant. While this specification potentially accounts for the factor structure, the results remain questionable due to the strong cross-section dependence. To overcome this weakness, in column (5), we then employ the CCE estimator as per equation \eqref{eq:CCEModel}---stage (1) of our three step procedure---and add, in addition to the observed common factors, cross-section averages to fully account for the unobserved factor structure.\footnote{We add the cross-section averages of the dependent variable, real exchange rate, GDP growth, the real interest rate, capital-to-asset ratio and money over GDP. We do not add the cross-section averages of the macroprudential variables as they are part of the observed common factor 'World Macroprudential'. The six cross-section averages ensure that the rank condition holds, i.e., the number of (un)observed factors is smaller or equal to the number of cross-section averages, see \cite{Karabiyik2017,Vos2024}. Together with the observed common factors, the cross-sectional averages span variables that are under direct policy control (like the exchange rate choice or macroprudential regulations) and those that are not under direct control (e.g., growth).}\(^,\)\footnote{We also note that, in a very general setting, policy controlled variables in \(x_{i,t}\) can interact with the common factors, resulting in higher order functions. In practice, these non-linear factors are linear in parameters and can be treated as additional factors in the estimation. An advantage of the CCE estimator is that we do not require knowledge of the exact factor structure. In the case of a time-invariant policy variable, such as the fixed or flexible exchange rate regime, the heterogeneous factor loadings soak up the interaction with the common factors. The same holds for interactions of the principal components and time-invariant policy variables when using a heterogeneous slope estimator. When using a pooled estimator, the interaction becomes part of the error term, resulting in strong cross-sectional dependence and biased estimates of the coefficients. }
The CD-test statistic falls into a non-rejection area (CD=0.797) and the estimate of the number of common factors in the residual is zero so that we can now calculate the composite \(\hat{u}_{i,t}\), which encapsulates the entire factor structure. 
As we found three common factors in the non-core to core ratio, in the second step, we extract the first three principal components (PCs) from the composite \(\hat{u}_{i,t}\). The PCs account for roughly 80\% of the explained variance, see the lower panel in Table  \ref{tab:MainResultsAdd}.\footnote{We note that the estimate for the number of factors in the composite, \(\hat{M}_{\hat{u}}\), is 2, the same as the number of factors in the residuals of the (TW)FE regressions. No factors remain in the residual, so we conclude that the composite captures the entire factor structure.}
Columns (6) and (7) of Table \ref{tab:MainResults} present the results of the 3rd stage. Introducing the PCs successfully mitigates cross-sectional dependence, as indicated by the CD-statistic of -0.134 (p=0.894). As explained in the previous section, because the estimated PCs span both observed and unobserved global factors, their inclusion in the MG regression makes the observed common factors redundant (and they display high multicollinearity with some of them, as discussed below), so the latter are omitted.  
Hence we conclude that these two specifications fully account for the factor structure. In column (6), all three PCs are significant and positive, implying an increase in the non-core ratio following a shock propagated by the PCs. Column (7) then tests whether global factors (captured by the three PCs) affect the non-core ratio of countries on a fixed exchange rate more than those on a flexible exchange rate. The loadings for the first PC is higher in countries subject to fixed exchange rates (0.984 vs. 0.302) and significant. PC2 is positive and significant for fixers, while PC3 is significant for floaters. In general the loadings for fixers are larger than those for floaters, as confirmed by the test statistic of 7.12 on the differences between the sum of the PC loadings between fixed and flexible exchange rate regimes, shown in the column labelled (7) in Table \ref{tab:MainResultsAdd}. Therefore, the upshot is that there is cause for the trilemma hypothesis applied to banks' non-core funding ratios.

Across all specifications, it is mainly three domestic variables that affect banks' non-core ratios in a statistically significant manner. First, the lagged dependent variable is positive and significant at the 1\% level throughout. Second, banking sectors in countries with a more appreciated real exchange rate have higher non-core ratios. This result might reflect that more an appreciated real exchange rate makes it chaeper to borrow abroad. Finally, higher money-to-GDP ratios also tend to increase banks' non-core ratios, consistent with the evidence on drivers of credit booms and the role of cross-border borrowing found in the studies referred to in previous sections.

\subsection{Robustness Checks}
\label{robsec}

In this sub-section, we show that our main results corresponding to column (7) of Table \ref{tab:MainResults} are robust along several dimensions. We first drop the United States from our sample, as most of the observed common factors included in the regressions are based on (or at least dominated by) the US. Our benchmark results are hardly affected and we still find PC1 and PC3 to affect floaters, and PC1 and PC2 to affect fixers to a statistically significant extent, as column (1) of Table \ref{tab:rob} shows.

In column (2), we use our benchmark sample including Iceland, which we dropped for the benchmark analysis because it is an extreme outlier for several of the variables included in the analysis. As becomes apparent, most results are similar to our baseline results, but the coefficients are estimated more imprecisely, at least for floaters, presumably due to the additional uncertainty introduced by the inclusion of Iceland. In addition, the specification is left with some cross-sectional dependence.

In column (3), we drop the Covid pandemic and end the sample period in 2019:Q4. This change has virtually no effects on our main results, despite a slight drop in statistical significance. Column (4) additionally starts the sample in 2010:Q1 in order to eliminate potential effects of the global financial crisis and to focus on an entire 'no-crisis' sample of 2010:Q1-2019:Q4. The results show that, as in the benchmark analysis, PC1 and PC2 are larger for countries with a fixed exchange rate. Yet, PC3 is now insignificant for both country groups. As we argue in Section \ref{dissecting} below, PC3 mostly captures financial crisis effects, so it is no surprise to see that PC3 turns insignificant in a 'no-crisis' sample. We also note that the difference in the loadings of fixers vs. floaters to the first principal component is now substantially larger than in our benchmark regression, suggesting that exchange rate insulation is stronger outside of major global crisis periods.

Finally, in column (5), we use an alternative outcome variable---banks' loan-to-deposit ratio. Previous work suggested it to be closely correlated with the non-core to core ratio, as we also confirm for our sample in Section 2. For this outcome variable, we find PC1 and PC2 to be significant drivers in countries with a flexible exchange rate and PC1 only to be statistically significant for countries on a fixed exchange rate. One difference to the previous results that we obtained for banks' non-core ratios is that a flexible exchange rate does not provide any insulation to PC1. This result, however, is driven by the crises episodes included in column (5). In fact, when in unreported specifications we drop the Covid episode and/or the global financial crisis from the sample, we find that the loadings to PC1 are substantially larger for fixers than for floaters, as in our benchmark analysis.

\begin{table}%
%\centering
\tiny
\resizebox{\textwidth}{!}{\begin{tabular}{l cc cc c  }\hline\hline
                              &         (1)   &         (2)   &         (3)   &         (4)   &         (5)   \\
                              &       Excl.   &       Incl.   &          No   &          No   &LTD \\ & US & Iceland & COVID & Crisis &    \\
\hline
Lag Dep. Var.              &       0.652   &       0.669   &       0.609   &       0.422   &       0.694   \\
                              &     (0.030)***&     (0.034)***&     (0.035)***&     (0.046)***&     (0.038)***\\
GDP Growth                     &      -0.021   &       0.051   &      -0.044   &       0.075   &      -0.070   \\
                              &     (0.104)   &     (0.115)   &     (0.096)   &     (0.069)   &     (0.060)   \\
Real Interest Rate                    &      -0.093   &      -0.088   &      -0.094   &       0.066   &      -0.086   \\
                              &     (0.074)   &     (0.100)   &     (0.139)   &     (0.099)   &     (0.059)   \\
REER                      &       0.413   &       0.509   &       0.344   &      -0.155   &      -0.031   \\
                              &     (0.185)** &     (0.194)***&     (0.147)** &     (0.229)   &     (0.071)   \\
Capital-to-Asset             &       0.837   &       0.977   &       0.755   &       1.344   &       0.276   \\
                              &     (0.498)*  &     (0.550)*  &     (0.428)*  &     (0.875)   &     (0.387)   \\
Capital Reg.                &      -4.174   &      -2.513   &      -5.207   &      -1.285   &       0.396   \\
                              &     (3.005)   &     (2.333)   &     (3.451)   &     (1.875)   &     (0.776)   \\
Liquidity Reg.                 &       0.207   &      -0.212   &       2.400   &      -2.474   &       1.420   \\
                              &     (3.332)   &     (2.609)   &     (5.279)   &     (3.556)   &     (1.250)   \\
Money / GDP                 &       0.206   &       0.122   &       0.172   &       0.508   &      -0.036   \\
                              &     (0.065)***&     (0.082)   &     (0.142)   &     (0.443)   &     (0.089)   \\
\multicolumn{5}{l}{Principal Components (PC, fixed = 0)}\\ \ PC1                        &       0.307   &       0.241   &       0.326   &       0.413   &       0.680   \\
                              &     (0.164)*  &     (0.139)*  &     (0.187)*  &     (0.165)** &     (0.206)***\\
\ PC2                        &      -0.091   &       0.767   &      -0.120   &       0.013   &       0.227   \\
                              &     (0.155)   &     (0.723)   &     (0.188)   &     (0.154)   &     (0.124)*  \\
\ PC3                        &       0.407   &       1.416   &       0.435   &       0.184   &      -0.086   \\
                              &     (0.218)*  &     (1.201)   &     (0.224)*  &     (0.156)   &     (0.237)   \\
\multicolumn{5}{l}{Principal Components (PC, fixed = 1) }\\ \ PC1                        &       1.175   &       0.962   &       1.426   &       1.667   &       0.394   \\
                              &     (0.586)** &     (0.452)** &     (0.737)*  &     (0.812)** &     (0.238)*  \\
\ PC2                        &       0.760   &      -0.415   &       0.634   &       0.883   &       0.418   \\
                              &     (0.452)*  &     (0.248)*  &     (0.581)   &     (0.386)** &     (0.256)   \\
\ PC3                        &       0.844   &       0.747   &       0.741   &       0.856   &       0.275   \\
                              &     (0.585)   &     (0.444)*  &     (0.591)   &     (0.673)   &     (0.167)   \\
\hline  Obs                   &        2088   &        2232   &        1890   &        1200   &        2160   \\
Countries                     &          29   &          31   &          30   &          30   &          30   \\
Time Periods                  &          72   &          72   &          63   &          40   &          72   \\
\hline CD                     &      -0.415   &       2.869   &      -1.858   &       0.085   &       0.547   \\
\ p                           &       0.678   &       0.004   &       0.063   &       0.932   &       0.584   \\
\(\hat{M}_{\hat{\epsilon}}\)&           0   &           0   &           0   &           0   &           0   \\
\hline \(R^2\)                &       0.042   &       0.092   &       0.043   &       0.036   &       0.021   \\
\(R^2\) (MG)                  &       0.947   &       0.885   &       0.944   &       0.944   &       0.973   \\
\(R^2\) (within)              &       0.958   &       0.908   &       0.957   &       0.964   &       0.979   \\
\hline\multicolumn{5}{l}{1st Stage} \\ CD&       0.538   &       3.430   &      -0.558   &      -0.085   &      -0.306   \\
\ \ p                         &       0.591   &       0.001   &       0.577   &       0.933   &       0.760   \\
\(\hat{M}_{\hat{u}}\)     &           2   &           1   &           1   &           1   &           2   \\
\(\hat{M}_{\hat{\epsilon}}\)&           0   &           0   &           0   &           0   &           0   \\\hline\hline

\end{tabular}}
\justify{This table displays the robustness checks. The non-core ratio is the dependent variable in Columns (1)-(4). In Column (5), \textit{LTD}, the loan-to-deposit ratio, is the dependent variable. Standard errors in parenthesis. Significance levels: \(^* 10\%, ^{**}5\%, ^{***}1\%\). 
Column \textit{Excl. US} excludes the United States. Column \textit{Inc. Iceland} includes Iceland. Column \textit{No Covid} spans the time period 2004:Q1-2019:Q4 and \textit{No Crisis} 2010:Q1-2019:Q4. The 1st stage for each column is not displayed. The PCs are obtained from this initial stage. \(\hat{M}_{X}\) is the estimate of the number of common factors using the Growth Criterion from \cite{AhnHorenstein2013}. \(\hat{u}\) is the composite of observed and unobserved common factors from Equation \eqref{eq:uhat} and \(\hat{\epsilon}\) the residual. All variables are defined as in Table \ref{datasources}. For further details see notes of Table \ref{tab:MainResults}.}
 \caption{\scshape Robustness Checks}
 \label{tab:rob}
\end{table}

\subsection{Does Macroprudential Policy Insulate Countries from the GFC?}

\begin{table}%
%\centering
\tiny
\begin{tabular}{l cc c  }\hline\hline
                              &         (1)   &         (2)   &         (3)   \\
                              &         All   &   Fixed = 0   &   Fixed = 1   \\
\hline
Lag Dep. Var.              &       0.659   &       0.701   &       0.623   \\
                              &     (0.031)***&     (0.045)***&     (0.041)***\\
GDP Growth                     &      -0.013   &      -0.001   &      -0.023   \\
                              &     (0.091)   &     (0.042)   &     (0.169)   \\
Real Interest Rate                    &      -0.065   &       0.125   &      -0.232   \\
                              &     (0.072)   &     (0.070)*  &     (0.104)** \\
REER                      &       0.424   &       0.079   &       0.726   \\
                              &     (0.186)** &     (0.036)** &     (0.334)** \\
Capital-to-Asset             &       0.807   &       0.550   &       1.031   \\
                              &     (0.482)*  &     (0.609)   &     (0.744)   \\
Capital Reg.                &      -4.051   &      -0.280   &      -7.351   \\
                              &     (2.908)   &     (0.826)   &     (5.346)   \\
Liquidity Reg.                 &      -0.305   &      -0.128   &      -0.461   \\
                              &     (2.977)   &     (0.959)   &     (5.606)   \\
Money / GDP                 &       0.176   &       0.081   &       0.259   \\
                              &     (0.064)***&     (0.053)   &     (0.108)** \\
\multicolumn{4}{l}{Principal Components (Macro. Tight. = 0)}\\ \ PC1                        &       0.405   &       0.135   &       0.640   \\
                              &     (0.458)   &     (0.269)   &     (0.835)   \\
\ PC2                        &       0.792   &       0.020   &       1.468   \\
                              &     (0.311)** &     (0.097)   &     (0.527)***\\
\ PC3                        &       0.935   &       0.228   &       1.554   \\
                              &     (0.550)*  &     (0.178)   &     (1.008)   \\
\multicolumn{4}{l}{Principal Components (Macro. Tight. = 1) }\\ \ PC1                        &       0.881   &       0.511   &       1.204   \\
                              &     (0.304)***&     (0.220)** &     (0.533)** \\
\ PC2                        &      -0.083   &      -0.192   &       0.012   \\
                              &     (0.243)   &     (0.323)   &     (0.366)   \\
\ PC3                        &       0.259   &       0.571   &      -0.014   \\
                              &     (0.291)   &     (0.401)   &     (0.418)   \\
\hline  Obs                   &        2160   &        1008   &        1152   \\
Countries                     &          30   &          14   &          16   \\
Time Periods                  &          72   &          72   &          72   \\
\hline CD                     &      -0.134   &       2.126   &       0.905   \\
\ p                           &       0.894   &       0.033   &       0.365   \\
\(\hat{M}_{\hat{\epsilon}}\)&           0   &           0   &           0   \\
\hline \(R^2\)                &       0.042   &       0.057   &       0.041   \\
\(R^2\) (MG)                  &       0.947   &       0.929   &       0.949   \\
\(R^2\) (within)              &       0.958   &       0.943   &       0.959   \\
\hline\multicolumn{4}{l}{1st Stage} \\ CD&       0.797   &       0.797   &       0.797   \\
\ \ p                         &       0.425   &       0.425   &       0.425   \\
\(\hat{M}_{\hat{u}}\)     &           2   &           2   &           2   \\
\(\hat{M}_{\hat{\epsilon}}\)&           0   &           0   &           0   \\\hline\hline

\end{tabular}
\justify{Principal Components obtained from Baseline Regression in Table \ref{tab:MainResults}. The regressions in this table identify the loadings to the PCs separately for countries with loose vs. tight  macroprudential policies. Column \textit{Fixed = 0} restricts the sample to countries with flexible exchange rates, \textit{Fixed = 1} to countries with fixed exchange rates. See the additional notes of Table \ref{tab:MainResults}. All variables are defined as in Table \ref{datasources}.  For further details see notes of Tables \ref{tab:MainResults} and \ref{tab:rob}.}
 \caption{\scshape The Relevance of Macroprudential Policies}
 \label{tab:Macro}
\end{table}

Previous regressions have shown that a flexible exchange rate helps insulate countries from the effects of the global financial cycle. In this sub-section, we examine whether an important policy variable---the stance of macroprudential regulation---has a similar effect. To this end, we estimate a version of column (7) in Table \ref{tab:MainResults} where we separate the loadings to the PCs according to a country's macroprudential stringency, and not its exchange rate regime. Specifically, we define a country as having tighter macroprudential regulations when it, cumulatively during our sample period, tightened its macroprudential policy stance more than the median country. As column (1) of Table \ref{tab:Macro} shows, the loadings to the second and third PC are substantially smaller and statistically insignificant for countries with tighter macroprudential regulation. For PC1, we find the loadings to be larger for countries with tighter macroprudential regualtions. As we show below, PC1 correlates strongly with the worldwide macroprudential strength, so that it is natural to see that PC1 loads more heavily on countries that follow the world trend more closely, i.e., have a tighter regulation. In columns (2)-(3), we split the overall sample into countries subject to a flexible and fixed exchange rate, respectively, and obtain a similar picture, but the statistical significance declines with the smaller sample size. 

Interestingly, we find that having a stricter macroprudential policy stance especially helps countries subject to fixed exchange rates insulate against PC2 and PC3, which together explain about 21\% of variations in non-core ratios. For instance, the loading to PC2 (PC3) is 1.47 (1.55) for fixers with loose macroprudential regulation, whereas it is essentially zero for fixers with a tighter regulation. For countries on a flexible exchange rate, macroprudential tightness does not provide any additional insulation device.

\section{Inspecting the Mechanisms}
\label{dissecting}

In this section, we look at the mechanisms through which the global financial cycle affects banks' non-core ratios. To this end, we first correlate our principal components with observable global factors to see which ones are the most significant drivers of our PCs. This is important for policy makers that typically monitor such observed global factors carefully. Afterwards, we analyze whether the changes in non-core ratios documented above are driven by domestic- vs. foreign-currency borrowing of banks, and whether the foreign funds are provided by other banks via the interbank market, nonbank financials, or other sectors of the economy.

\subsection{Which Global Variables Correlate with our Principal Components?}

\begin{table}%
\centering
\tiny
\begin{tabular}{@{\extracolsep{4pt}}l cc cc cc  @{}}
	\hline\hline
& (1)& (2)& (3)& (4)& (5)& (6)\\\cline{2-3}\cline{4-5}\cline{6-7}
& \multicolumn{2}{c}{PC1} & \multicolumn{2}{c}{PC2} & \multicolumn{2}{c}{PC3} \\ \cline{2-3}\cline{4-5}\cline{6-7}
VIX                   &      -0.434   &      -0.484   &       0.015   &       0.027   &       0.142   &      -0.180   \\
                              &     (0.153)***&     (0.171)***&     (0.135)   &     (0.126)   &     (0.172)   &     (0.184)   \\
Oil Price              &      -0.404   &      -0.334   &      -0.142   &      -0.125   &       0.214   &       0.414   \\
                              &     (0.223)*  &     (0.228)   &     (0.191)   &     (0.192)   &     (0.224)   &     (0.211)*  \\
Shadow Rate            &       1.094   &       1.042   &       1.171   &       1.168   &       1.076   &       0.852   \\
                              &     (0.098)***&     (0.101)***&     (0.096)***&     (0.094)***&     (0.109)***&     (0.114)***\\
Non-Bank Share          &      -0.777   &      -0.601   &       0.639   &       0.683   &      -1.046   &      -0.546   \\
                              &     (0.237)***&     (0.247)** &     (0.224)***&     (0.236)***&     (0.281)***&     (0.261)** \\
World GDP Growth           &       0.215   &       0.200   &      -0.271   &      -0.275   &       0.177   &       0.137   \\
                              &     (0.124)*  &     (0.116)*  &     (0.086)***&     (0.081)***&     (0.134)   &     (0.112)   \\
World Macropru.           &      -2.957   &      -3.023   &      -0.580   &      -0.584   &       1.720   &       1.435   \\
                              &     (0.205)***&     (0.209)***&     (0.190)***&     (0.194)***&     (0.210)***&     (0.185)***\\
World Inflation        &       0.068   &       0.030   &       0.347   &       0.338   &      -0.002   &      -0.110   \\
                              &     (0.126)   &     (0.130)   &     (0.080)***&     (0.082)***&     (0.107)   &     (0.106)   \\
US REER               &      -0.778   &      -0.640   &       0.804   &       0.830   &      -0.710   &      -0.255   \\
                              &     (0.258)***&     (0.257)** &     (0.241)***&     (0.232)***&     (0.289)** &     (0.284)   \\
GFC                           &               &       0.566   &               &       0.083   &               &       2.050   \\
                              &               &     (0.290)*  &               &     (0.235)   &               &     (0.456)***\\
COVID                         &               &      -0.331   &               &      -0.225   &               &       0.114   \\
                              &               &     (0.625)   &               &     (0.384)   &               &     (0.579)   \\
\hline  Time Periods          &          72   &          72   &          72   &          72   &          72   &          72   \\
\(R^2\)                       &       0.971   &       0.973   &       0.921   &       0.921   &       0.675   &       0.787   \\\hline\hline

\end{tabular}
\justify{Regressions of the PCs on observed common factors, with PCs obtained from Table \ref{tab:MainResults} and all regressors being standardized. Global Financial Crisis (GFC) and COVID are dummies equal to one for 2007:Q4-2011:Q4, following \cite{laeven2018systemic}, and 2019:Q4-2020:Q4, respectively. All variables are defined as in Table \ref{datasources}. }
 \caption{\scshape Regressions of our Principal Components on  Observed Common Factors}
\label{tab:dissecting}
\end{table}

A possible reservation to our approach is the reliance of our estimation on PCs that possibly span a variety of observed and unobserved global factors, implying that those PCs lack a straightforward economic meaning; this in turn would make the results less valuable from a policy-intervention standpoint.

We address this concern here by relating the PCs to observable common factors that have been featured in previous studies and also in the earlier specifications in Tables \ref{tab:MainResults}. This is done in Table \ref{tab:dissecting}, which reports regressions of each PC on that set of observed global factors, with the latter being standardized by subtracting the mean and dividing by the standard deviation of each variable. 

As becomes apparent, the first PC loads by far most significantly on the macropudential stringency variable (column 1). In particular, tighter macroprudential regulations reduce banks' non-core ratios. The same is true for a higher VIX, higher oil prices, a higher world non-bank share, a lower US shadow rate and a stronger USD, but their individual effects on the PC and hence banks' non-core ratios are smaller. Column (3) of Table \ref{tab:dissecting} shows that PC2 is mostly driven by the US shadow rate, with a correlation between both variables being positive. 
At first sight, this positive relation between US rates and banks' foreign borrowing seems inconsistent with the previous global financial cycle literature (e.g., \citealp{miranda2020us}), which shows that \textit{lower} US rates increase bank leverage and cross-border capital flows. Note, however, that changes in the US shadow rate during our sample period are predominantly driven by the Fed's quantitative easing and tightening cycles. Larger central bank purchases of long-term bonds (lower shadow rates) reduce long-term yields and flatten the yield curve (\citealp{d2013flow}; \citealp{sudo2021quantifying}). This, in turn, should then incentivize banks to fund any lending expansions with long-term bond issuances and not with short-term borrowing via the interbank market. 
As we show in Section \ref{bis}, the changes in non-core ratios we observe are mostly driven by changes in interbank borrowing, so it is not a surprise to see that lower US shadow rates tend to reduce banks' non-core ratios. In fact, when we use the loan-to-deposit ratio as outcome variable and correlate the PCs extracted from this specification with the US shadow rate, we obtain the same negative coefficient as the extant literature (see Appendix Table \ref{tab:Robust_LTD_PCRegs}). The same is true when we correlate the US shadow rate and PC2 extracted from our non-core ratio for a sample period that excludes any quantitative easing or tightening policies. For instance, before 2007, the correlation between both variables is -15.7\%.

Finally, column (5) of Table \ref{tab:dissecting} shows that PC3 also correlates most significantly with macroprudential policies, but the coefficient estimate is smaller than that in column (1) and the \(R^2\) of this regression is quite small (0.68). A graphical inspection of this PC in Figure \ref{pcfig} shows that PC3 tends to increase during crises. We hence add to the previous specification two crisis dummies---one US banking crisis indicator that is one between 2007:Q4 and 2011:Q4\footnote{\citet{laeven2018systemic} identify a US banking crisis for this period.} and one Covid dummy that takes a value of one for the initial phase of the Covid pandemic, i.e., between 2019:Q4 and 2020:Q4. For the sake of completeness, we also add the same crisis dummies to the regressions of PC1 and PC2. Adding the crisis dummies significantly increases the \(R^2\) for the PC3 regression (column 6). The US banking crisis dummy further turns out being economically most significant and clearly dominates the macroprudential stance variable. In contrast, the crisis dummies do not affect PC1 and PC2 to the same extent as PC3 (columns 2 and 4). 

We have shown before that the loadings to PC1 and PC2 are stronger for countries with a fixed exchange rate, whereas the effect of PC3 is quite similar in economic terms for fixers and floaters. Therefore, having a flexible exchange rate helps insulating against influences stemming from regulatory, financial or real variables, all of which are important drivers of PC1 and PC2, whereas it does not help insulating against the effects of financial crises, which seem to drive PC3. This novel result is consistent with column (4) of Table \ref{tab:rob}.

To summarize, we establish that PC1 is driven mostly by changes in the macroprudential policy stance, PC2 by the US shadow rate, and PC3 by crisis dummies. The tight relation between our PCs and these observed common factors can also be seen graphically in Figures \ref{fig:pc1}-\ref{fig:pc3}. In contrast, our results presented in this sub-section suggest that the VIX, oil prices, and the US real exchange rate---variables that have been highlighted in previous work on the GFC---carry the expected sign but generally have a smaller impact on banks' non-core ratio. The short of this is that there is evidence that global observables affect non-core ratios in a way that involves complex interactions among real, monetary and financial variables that are best summarized by a handful of composite indicators in the form of estimated principal components. This does not need to be invaluable to policy makers to the extent that we know that each of these PCs correlates tightly with those observables that are “usual suspects” when it comes to monitoring crisis risk and designing policy interventions. Our analysis brings home the important point that---to the extent that fluctuations in the non-core ratio are related to macro risks---those risks are best seen as a composite of the inter-play of the observed and unobserved global factors.

\subsection{A Breakdown of Banks' Foreign Liabilities}
\label{bis}

In previous estimations, we show that our principal components positively affect banks' non-core ratios, which we defined as banks' foreign liabilities over total deposits. One disadvantage of the IFS data, however, is that they do not allow for a detailed breakdown into different forms of cross-border bank liabilities. We fill this gap by exploiting data on foreign liabilities from the BIS Locational Banking Statistics, which provides a breakdown into (i) domestic- and foreign-currency borrowing of banks, and (ii) foreign liabilities vis-a-vis banks, nonbank financials and all other sectors for 23 of the 30 countries included in our benchmark sample. We then relate these different components of banks' foreign liabilities in standard fixed effects regressions to the principal components extracted above to see at a more granular level how the global financial cycle affects banks' non-core funding ratios. %Given the summary statistics of Table \ref{datatable}, which show that most of banks' foreign borrowing comes (i) in the form of foreign-currency, not domestic-currency loans and (ii) is provided by other banks via the interbank market, and not by nonbank financials or other lenders, we hypothesize that the foreign-currency component and liabilities originating from other banks are mostly affected by our PCs.

\begin{table}[h]
\centering
\tiny
\begin{tabular}{l cc cc cc  }
	\hline\hline
            &         (1)   &         (2)   &         (3)   &         (4)   &         (5)   &         (6)   \\
            &       Total   &     Foreign   &    Domestic   &       Banks   &   Non-Banks   &       Other   \\
\hline Lag Dep. Var.      &       0.958   &       0.957   &       0.950   &       0.954   &       0.961   &       0.951   \\
            &     (0.015)***&     (0.015)***&     (0.018)***&     (0.019)***&     (0.013)***&     (0.015)***\\
PC1         &       0.197   &       0.100   &       0.110   &       0.199   &       0.017   &      -0.050   \\
            &     (0.067)***&     (0.032)***&     (0.043)** &     (0.061)***&     (0.017)   &     (0.023)** \\
PC2         &       0.250   &       0.106   &       0.140   &       0.218   &       0.049   &      -0.106   \\
            &     (0.127)** &     (0.058)*  &     (0.089)   &     (0.110)** &     (0.039)   &     (0.041)***\\
PC3         &       0.395   &       0.128   &       0.279   &       0.388   &       0.014   &       0.043   \\
            &     (0.221)*  &     (0.097)   &     (0.147)*  &     (0.175)** &     (0.058)   &     (0.062)   \\
\hline Obs  &        1633   &        1627   &        1627   &        1554   &        1633   &        1554   \\
\(R^2\)     &       0.981   &       0.981   &       0.977   &       0.976   &       0.981   &       0.963   \\ \hline\hline

\end{tabular}
\justify{In this table, we regress in standard fixed effects regressions several BIS cross-border variables (scaled by deposits) on the principal components extracted from \(\hat{u}_{i,t}\) in our baseline regression. Column (1) uses total cross-border liabilities, columns (2) and (3) break them down into foreign-currency and domestic-currency liabilities and columns (4)-(6) into liabilities provided by banks, nonbank financials and all other sectors of the economy.}
\caption{BIS Variables Regressed on Principal Components}
\label{tab:BIS}
\end{table}

Table \ref{tab:BIS} contains the attendant results which show that all three principal components are positively correlated with banks' \textit{total} cross-border BIS liabilities, consistent with our benchmark results (column 1). The breakdown into the different components, shown in columns (2)-(6), indicate that this increase materializes via both increases in domestic- and foreign-currency liabilities, as well as an increase in banks' liabilities vis-a-vis banks. In contrast, liabilities vis-a-vis other financials or all other sectors are hardly (in some cases even negatively) correlated with our principal components. As Table \ref{tab:BIS_Interaction} of the Appendix shows, at least some of these effects are again stronger for countries on a fixed exchange rate. Overall, this is evidence that the relation between the global financial cycle and non-core ratios documented in this paper works via an increase in banks' foreign funding provided via the interbank market---a funding source that is well-known to be quite flighty. The currency denomination does not seem to matter a lot in our sample.\footnote{This may or may not hold for a broader sample that includes emerging markets where currency mismatches between banks' assets and liabilities are much larger and exchange rates are far more volatile than in our sample of advanced countries. So, we acknowledge that this result may be somewhat sample specific.}

\section{Conclusion}
\label{conclusion}

This paper has employed quarterly country-level data over the period 2004-2022 to study the relationship between the global financial cycle and the foreign  borrowing component of the non-core funding ratio of advanced economies’ banking systems. The focus on such a non-core ratio is grounded in the mechanism highlighted in earlier work: as bank capital is sticky and deposits are not too elastic to economic growth, rapid increases in bank loans have to be funded via the so-called non-core liabilities, of which foreign borrowing has been shown to be a key component. Such a non-core ratio should then be tightly correlated with the ratio of bank loan-to-deposits and leverage in general, which we know are important drivers of the business cycle since the seminal work of \citet{bernanke1988money} and \citet{bernanke1999financial} on the credit channel and macroeconomic fluctuations. By providing new econometric evidence on the variety and form of which global factors drive banks’ non-core ratios, this paper broadens earlier findings on how global factors are powerful drivers of bank lending and hence aggregate fluctuations in advanced economies through the bank funding channel.

Further, our econometric exercise indicates that most country-specific factors only play a minor role in affecting banks' non-core ratios, consistent with the fact that much of banking system in advanced countries is highly globalized---including through the role of global banks in shaping leverage also in local banks. At the same time, we also find that global factors driving non-core funding ratios cannot be spanned by a single variable like the VIX or any other financial or real factor. Therefore, the global financial cycle might be best represented by composites of global factors, i.e., global principal components. 

A key finding is that flexible exchange rates play a significant insulation role in the way global factors affect the ratio of non-core to core funding. In particular, global shocks affect non-core ratios less for countries on a flexible exchange rate regime and even more strongly outside of global crises episodes, like the GFC or the Covid pandemic. We also document that tighter macroprudential regulations help insulate bank funding from the global financial cycle.

%Our findings highlight an additional channel through which the Mundelling trilemma  remains alive, thereby also stressing the need for due attention to be paid to banking sector-macro connections when discussing monetary and macroprudential policy trade-offs.

\newpage
\singlespacing
\bibliography{main}
\bibliographystyle{aer}

\newpage
\appendix

\addcontentsline{toc}{section}{Appendix}

\numberwithin{subsection}{section}

%\numberwithin{table}{section}

%\numberwithin{figure}{section}

\numberwithin{equation}{section}

\renewcommand*\theequation{A\arabic{equation}}

% Remove "A" from table names
\renewcommand{\thetable}{\arabic{table}}
\renewcommand{\thefigure}{\arabic{figure}}

\setcounter{equation}{0}

\section{Data Appendix}
\label{data_details} 

In this appendix, we describe in detail how we constructed our country-level panel data set. Table \ref{countrylist} contains the countries included in the analysis, while Table \ref{datasources} shows an overview of the variables employed in the analysis. 

The starting point of our data set is the non-core and loan-to-deposit ratio, which we source from the IMF's IFS database for countries that the World Bank classifies as high-income. We then merge them with several global variables and domestic controls, most of which we obtain from the IFS database and the Federal Reserve Bank of St. Louis (Fred). Note that our estimations require a balanced panel, so we need to ensure that missing data points are filled with information from web sources and other databases, or are extrapolated using the procedures outlined below.

\bigskip 

\textbf{Country-Specific Variables}

\bigskip

\underline{Non-Core Ratio}: We compute this variable as IFS line 26c (foreign liabilities) over the sum of lines 24 (transferable deposits included in broad money) and 25 (other deposits included in broad money). Note that, since all of the IFS variables are missing for a few country-quarter pairs, we interpolate missing values linearly, in line with \cite{miranda2020us}. In addition, as the coverage of two important advanced economies---the UK and Canada---in the IFS database is limited, we hand-match banking sector balance sheet data from the Bank of Canada and Bank of England.

\underline{Loan-to-Deposit Ratio}: We compute this variable as IFS line 22d (claims on private sector) over the sum of lines 24 (transferable deposits included in broad money) and 25 (other deposits included in broad money). The note above on the UK and Canada applies here as well. 

\underline{Capital-to-Asset Ratio}: We compute this variable as IFS line 27a (shares and other equity) over total assets. The note above on the UK and Canada applies here as well. For Canada, the Bank of Canada provides no capital data after 2020:Q3. We hence match Canadian capital ratios from the Financial Soundness Indicators (defined as capital and reserves over total assets). To avoid jumps in the series when using two different capitalization ratios, we first extrapolate one with the other and then replace missing data points for Canada after 2020:Q3 with this extrapolated value. For the euro area countries, where capital ratios are entirely missing in the IFS as well, we either use the corresponding series from the Financial Soundness Indicators as well, or---when unavailable for a given country-quarter pair---handmatch their capital ratios using the Fred database. As the latter only reports annual capital ratios, we interpolate the data linearly to obtain a quarterly series. Finally, for few countries, the series is not available until the end of the sample period. We hence replace the missing observations with the last available value, as capital ratios tend to be sticky over time.

\underline{BIS Cross-border Liabilities}: A country's cross-border bank liabilities based on the BIS Locational Banking Statistics, scaled by deposits (in USD to match the currency in the BIS database).

\underline{BIS Foreign-currency, cross-border Liabilities}: A country's foreign-currency cross-border bank liabilities based on the BIS Locational Banking Statistics, scaled by deposits (in USD to match the currency in the BIS database).

\underline{BIS Domestic-currency, cross-border Liabilities}: BIS Cross-border Liabilities net of BIS Foreign-currency, cross-border Liabilities.

\underline{BIS Cross-border Liabilities to Banks}: A country's cross-border bank liabilities vis-a-vis other banks based on the BIS Locational Banking Statistics, scaled by deposits (in USD to match the currency in the BIS database).

\underline{BIS Cross-border Liabilities to Nonbanks}: A country's cross-border bank liabilities vis-a-vis the nonbank financial sector based on the BIS Locational Banking Statistics, scaled by deposits (in USD to match the currency in the BIS database).

\underline{BIS Cross-border Liabilities to Other Sectors}: BIS Cross-border Liabilities net of BIS Cross-border Liabilities to Banks and net of BIS Cross-border Liabilities to Nonbanks.

\underline{GDP Growth}: Country-level, quarter-to-quarter change in the logarithm of real GDP. Source: IFS. When countries report the non-seasonally adjusted real GDP series, we use those; otherwise, we use seasonally-adjusted real GDP.

\underline{Nominal Interest Rate}: The nominal, short-term interest rate in \%. As default, we use the interbank interest rate. If it is not available for a specific country, we use the deposit interest rate or the 3-month treasury bill rate. Source: IFS. For countries where all three series are (mostly) missing, we match short-term interest rate data from the OECD. In a few cases, countries first report short-term interest rate data in the IFS database and then only in the OECD database. In this case, we first extrapolate one with the other and then replace missing data points with this extrapolated value in order to avoid jumps.

\underline{Inflation}: The quarter-to-quarter change in the logarithm of a country's CPI, in \%.

\underline{Real Interest Rate}: We define the real interest rate as the difference between the short-term nominal rate and the one-quarter lead of the quarter-to-quarter inflation rate (multiplied by 4 to obtain comparable magnitudes, as nominal rates are in \% per year).

\underline{REER}: A country's real effective exchange rate based on CPI. Main Source: IFS. In a few cases, the series in not available in the IFS and we fill the gaps using other sources, mainly the Fred database. We set 2010:Q1 as base quarter (index=100).

\underline{Money-to-GDP Ratio}: Broad money (M3) over nominal GDP (multiplied by 4 as GDP is quarterly), in \%. Main source: IFS. In several cases (mainly the euro area countries), the series in not available in the IFS and we fill the gaps using other sources, in particular the Fred database that also reports data on broad money for the euro area as a whole. In one case (Canada), we first have the money-to-GDP ratio in the IFS database and then only in the Fred database. In this case, we first extrapolate one with the other and then replace missing data points with this extrapolated value in order to avoid jumps in the series.

%\underline{Capital Account Openness}: The normalized Chinn-Ito index. Source: \citet{chinn2006matters}. The data come at the annual frequency and we interpolate them linearly to quarterly frequency. The data further end in 2021, and we assume that countries do not change their openness after that. Luxembourg is not covered by the \citet{chinn2006matters} database, but we set its capital account openness to the maximum of 1.

\underline{Liquidity Regulation}: The country-specific cumulative macroprudential liquidity regulation. Source: Based on the Liquidity variable of the IMF’s integrated Macroprudential Policy (iMaPP) Database, originally constructed by \citet{alam2019digging}. The original data come at monthly frequency and we transform it to quarterly frequency using simple averages.

\underline{Capital Regulation}: The country-specific cumulative macroprudential capital regulation. Source: Based on the sum of the CCB, Conservation, Capital and LVR variables of the IMF’s integrated Macroprudential Policy (iMaPP) Database, originally constructed by \citet{alam2019digging}. The original data come at monthly frequency and we transform it to quarterly frequency using simple averages.

\underline{Fixed}: =1 for countries with strictly fixed peg, i.e., countries where the coarse \citet{ilzetzki2019exchange} exchange rate classification is equal to 1. The data come at monthly frequency and we use end-of-quarter values for our quarterly data. The variable coverage ends in 2019, so we assume that countries do not switch their classification after that.\footnote{For our sample, the Czech Republic is the only country that switches its exchange rate regime (from flexible to fixed in 2014). We define the country as a floater throughout, but results are similar if we don't do so.}

\underline{Macroprudential Tightness}: =1 for countries that cumulatively during our sample period tightened their macroprudential policy stance more than the median country, based on the Sum17 variable of the IMF’s integrated Macroprudential Policy (iMaPP) Database, originally constructed by \citet{alam2019digging}.

\bigskip

{\vspace{20pt}}

\textbf{Global Variables}

\bigskip

\underline{VIX}: The log of the CBOE Volatility index. Source: Fred. 

\underline{REER US}: The US Real Effective Exchange Rate. Source: Fred.

\underline{Shadow Rate}: The US shadow interest rate. Source: \citet{wu2016measuring}. 

\underline{Oil Price}: WTI oil prices in USD. Source: Fred.

\underline{Non-Bank Share}: The share of financial intermediation by nonbanks. Source: Global Monitoring Report on Non-Bank Financial Intermediation, Financial Stability Board. The data come at the annual frequency and we interpolate them to quarterly frequency.

\underline{World GDP Growth}: The weighted average real GDP growth (quarter-to-quarter) of all countries classified as high-income by the World Bank. The applied weight is a country's nominal USD GDP.

\underline{World Inflation}: The weighted average inflation rate (quarter-to-quarter) of all countries classified as high-income by the World Bank. The applied weight is a country's nominal USD GDP.

\underline{World Macroprudential}: The weighted average cumulative macroprudential tightness variable of all countries classified as high-income by the World Bank. The applied weight is a country's nominal USD GDP. Source: Based on the Sum17 variable of the IMF’s integrated Macroprudential Policy (iMaPP) Database, originally constructed by \citet{alam2019digging}. The original data come at monthly frequency and we transform it to quarterly frequency using simple averages.

\bigskip

\begin{table}[h]
\centering

\caption{Countries Included in the Analysis and Their Insulation Characteristics}
\vspace{3mm}

\begin{tabular}{lrr}\hline\hline
Country & Fixed Exchange Rate & Macro Prudential Tightness\\ \hline
 Australia & 0 & 0 \\
Austria & 1 & 1 \\
Belgium & 1 & 0 \\
Canada & 0 & 1 \\
Chile & 0 & 0 \\
Cyprus & 1 & 0 \\
Czech Republic & 0 & 0 \\
Denmark & 1 & 1 \\
Estonia & 1 & 0 \\
Finland & 1 & 1 \\
France & 1 & 0 \\
Germany & 1 & 0 \\
Greece & 1 & 0 \\
Hungary & 0 & 1 \\
Iceland & 0 & 1 \\
Ireland & 1 & 0 \\
Israel & 0 & 1 \\
Italy & 1 & 0 \\
Japan & 0 & 0 \\
Korea, Republic of & 0 & 1 \\
Luxembourg & 1 & 1 \\
Netherlands & 1 & 1 \\
Norway & 0 & 1 \\
Poland & 0 & 1 \\
Portugal & 1 & 1 \\
Romania & 0 & 1 \\
Slovenia & 1 & 0 \\
Spain & 1 & 0 \\
Sweden & 0 & 1 \\
United Kingdom & 0 & 0 \\
United States & 0 & 0 \\ \hline

\end{tabular}

\label{countrylist}

\end{table}

\begin{landscape}

 		\begin{table}[tbh!]
 		\footnotesize
 		\centering
 		\begin{adjustbox}{max height=.95\hsize, max width=.95\textwidth}
\begin{tabular}{lrr}
 & Definition & Sources\tabularnewline
 &  & \tabularnewline
\hline 
\multicolumn{2}{l}{Bank Variables} & \tabularnewline
\hline 
 &  & \tabularnewline
Non-Core Ratio & Foreign liabilities over deposits & IFS, national CBs\\ 
Loan-to-Deposit Ratio & Loans to private sector over deposits & IFS, national CBs\\
Capital-to-Asset Ratio & Bank capital over total assets & IFS, Fred, FSI, national CBs\\
BIS Cross-border Liabilities & Cross-border bank liabilities over deposits & IFS, BIS LBS \\
BIS Foreign-currency, cross-border Liabilities & Cross-border bank liabilities denominated in foreign currency over deposits & IFS, BIS LBS \\
BIS Domestic-currency, cross-border Liabilities & Cross-border bank liabilities denominated in domestic currency over deposits & IFS, BIS LBS \\
BIS Cross-border Liabilities to Banks & Cross-border bank liabilities vis-a-vis banks over deposits & IFS, BIS LBS \\
BIS Cross-border Liabilities to Nonbanks & Cross-border bank liabilities vis-a-vis the nonbank financial sector over deposits & IFS, BIS LBS \\
BIS Cross-border Liabilities to Other Sectors & Cross-border bank liabilities vis-a-vis all other sectors (other than banks or nonbanks) over deposits & IFS, BIS LBS \\
 &  & \\
\hline 
\multicolumn{2}{l}{Domestic Controls} & \\
\hline 
 &  & \\
GDP Growth & Quarter-to-quarter real GDP growth rate & IFS \\
Real Interest Rate & The difference between the short-term nominal interest rate and the one-quarter lead inflation rate & IFS, OECD \\
REER & The real effective exchange rate based on CPI & IFS, Fred \\
Money-to-GDP Ratio & The broad money to GDP ratio & IFS, Fred \\
%Capital Account Openness & Capital account openness, normalized between 0 and 1 & \cite{chinn2006matters} \\
Liquidity Regulation & country-specific cumulative macroprudential liquidity
regulation & \cite{alam2019digging} \\
Capital Regulation & country-specific cumulative macroprudential capital
regulation & \cite{alam2019digging} \\
Fixed & Dummy=1 for strictly fixed exchange rate regimes (pegs) & \cite{ilzetzki2019exchange} \\
Macroprudential Tightness & Dummy=1 for countries with a cumulative macroprudential tightness stance above sample median & \cite{alam2019digging} \\
 & & \\
\hline 
\multicolumn{2}{l}{Global Variables} & \\
\hline 
 &  & \\
VIX & Logarithm of the implied volatility of S\&P 500 index options & Fred \\
REER US & The US real effective exchange rate & Fred \\
Shadow Rate & The US shadow interest rate & \citet{wu2016measuring} \\
Oil Price & WTI oil prices in USD & FRED \\
Nonbank Share & Worldwide share of financial intermediation by nonbanks & FSB \\
World GDP Growth & Current USD GDP weighted average quarter-to-quarter real GDP growth of high-income countries & IFS \\
World Inflation & Current USD GDP weighted average quarter-to-quarter inflation rate of high-income countries & IFS \\
World Macroprudential & Current USD GDP weighted average cumulative macroprudential tightness variable  & \cite{alam2019digging} \\ \hline

\end{tabular}
	\end{adjustbox}
 \caption{\scshape Variable Definitions and Sources}
 	\label{datasources}
 \end{table}
 %sideways
  
\end{landscape}

\clearpage
\newpage

\section{Additional Tables and Figures}
\label{data_apprendix}

\begin{table}[!h]
\scriptsize
\begin{tabular}{lcr}\hline\hline
	Sample & 2004-2022 & 2004-2019\\\hline
	Non-Core Ratio & 3 & 3 \\
	Loan-to-Deposit Ratio & 5 & 4 \\ 
	Observed Common Factor & 7 & 7 \\\hline\hline
\end{tabular}
\justify{
    Number of factors estimations based on \texttt{Growth Rate (GR)} estimator from \cite{AhnHorenstein2013}.  Variables in Observed common factors are: VIX, Oil Price, Shadow Rate, Non-Bank Share, World GDP Growth, World Inflation, World Macroprudential and the US real exchange rate. Estimations are done in \textit{Stata} using \texttt{xtnumfac} \citep{ReeseDitzen2023}.}    
    \caption{\scshape Estimated number of common factors}
\label{tab:NumComFac}
	\end{table}

\begin{table}[!h]
\centering
\tiny
\begin{tabular}{@{\extracolsep{4pt}}l cc cc cc  @{}}\hline\hline
& (1)& (2)& (3)& (4)& (5)& (6)\\\cline{2-3}\cline{4-5}\cline{6-7}
& \multicolumn{2}{c}{PC1} & \multicolumn{2}{c}{PC2} & \multicolumn{2}{c}{PC3} \\ \cline{2-3}\cline{4-5}\cline{6-7}
VIX                   &      -0.595   &      -0.657   &       0.204   &       0.005   &       1.033   &       0.871   \\
                              &     (0.199)***&     (0.257)** &     (0.228)   &     (0.241)   &     (0.196)***&     (0.226)***\\
Oil Price              &      -0.573   &      -0.490   &      -0.404   &      -0.261   &       0.331   &       0.405   \\
                              &     (0.252)** &     (0.272)*  &     (0.318)   &     (0.335)   &     (0.297)   &     (0.297)   \\
Shadow Rate            &       1.096   &       1.033   &      -0.566   &      -0.713   &       0.270   &       0.169   \\
                              &     (0.146)***&     (0.162)***&     (0.160)***&     (0.198)***&     (0.167)   &     (0.180)   \\
Non-Bank Share          &      -0.848   &      -0.637   &      -1.031   &      -0.672   &      -0.157   &       0.025   \\
                              &     (0.257)***&     (0.267)** &     (0.325)***&     (0.331)** &     (0.286)   &     (0.312)   \\
World GDP Growth           &       0.177   &       0.160   &       0.304   &       0.275   &      -0.069   &      -0.084   \\
                              &     (0.162)   &     (0.154)   &     (0.164)*  &     (0.152)*  &     (0.140)   &     (0.133)   \\
World Macropru.           &      -2.538   &      -2.619   &       1.601   &       1.414   &      -0.444   &      -0.573   \\
                              &     (0.260)***&     (0.277)***&     (0.274)***&     (0.290)***&     (0.273)   &     (0.287)*  \\
World Inflation        &      -0.125   &      -0.170   &      -0.195   &      -0.272   &       0.591   &       0.551   \\
                              &     (0.152)   &     (0.159)   &     (0.152)   &     (0.170)   &     (0.151)***&     (0.155)***\\
US REER               &      -1.174   &      -1.007   &      -1.699   &      -1.383   &       0.961   &       1.142   \\
                              &     (0.317)***&     (0.354)***&     (0.426)***&     (0.475)***&     (0.403)** &     (0.411)***\\
GFC                           &               &       0.686   &               &       1.396   &               &       0.855   \\
                              &               &     (0.420)   &               &     (0.542)** &               &     (0.379)** \\
COVID                         &               &      -0.381   &               &      -0.107   &               &       0.306   \\
                              &               &     (0.818)   &               &     (0.616)   &               &     (0.635)   \\
\hline  Time Periods          &          72   &          72   &          72   &          72   &          72   &          72   \\
\(R^2\)                       &       0.947   &       0.949   &       0.755   &       0.779   &       0.634   &       0.653   \\\hline\hline

\end{tabular}
\justify{Regressions of the PCs extracted from the loan-to-deposit regression in Table \ref{tab:rob} on the observed common factors.  Variables are standardized. Global Financial Crisis (GFC) and COVID are dummies equal to one for 2007:Q4-2011:Q4, following \cite{laeven2018systemic}, and 2019:Q4-2020:Q4, respectively. All variables are defined as in Table \ref{datasources}. }
 \caption{\scshape Regressions of Principal Components On the Observed Common Factors Extracted from Loan-to-Deposit Ratio Regressions}
\label{tab:Robust_LTD_PCRegs}
\end{table}

\begin{table}[!h]
\centering
\tiny
\begin{tabular}{l cc cc cc  }\hline\hline
                         &         (1)   &         (2)   &         (3)   &         (4)   &         (5)   &         (6)   \\
                         &       Total   &     Foreign   &    Domestic   &       Banks   &   Non-Banks   &       Other   \\
\hline Lag Dep. Var.                   &       0.954   &       0.952   &       0.948   &       0.949   &       0.960   &       0.950   \\
                         &     (0.016)***&     (0.015)***&     (0.018)***&     (0.021)***&     (0.013)***&     (0.016)***\\
\multicolumn{7}{l}{Principal Components (fixed = 0)}\\ PC1                     &       0.052   &       0.033   &       0.020   &       0.110   &      -0.007   &      -0.049   \\
                         &     (0.055)   &     (0.046)   &     (0.017)   &     (0.038)***&     (0.014)   &     (0.031)   \\
PC2                      &       0.003   &      -0.017   &       0.024   &       0.034   &       0.026   &      -0.058   \\
                         &     (0.084)   &     (0.067)   &     (0.033)   &     (0.073)   &     (0.025)   &     (0.031)*  \\
PC3                      &       0.256   &       0.237   &       0.035   &       0.306   &       0.031   &      -0.076   \\
                         &     (0.148)*  &     (0.119)** &     (0.049)   &     (0.114)***&     (0.041)   &     (0.059)   \\
\multicolumn{7}{l}{Principal Components (fixed = 1)}\\ PC1        &       0.257   &       0.122   &       0.155   &       0.195   &       0.041   &      -0.010   \\
                         &     (0.125)** &     (0.065)*  &     (0.072)** &     (0.100)*  &     (0.031)   &     (0.042)   \\
PC2        &       0.410   &       0.206   &       0.192   &       0.361   &       0.039   &      -0.096   \\
                         &     (0.217)*  &     (0.108)*  &     (0.149)   &     (0.220)   &     (0.068)   &     (0.077)   \\
PC3        &       0.240   &      -0.174   &       0.407   &       0.114   &      -0.027   &       0.211   \\
                         &     (0.373)   &     (0.181)   &     (0.243)*  &     (0.313)   &     (0.102)   &     (0.118)*  \\
\hline Obs               &        1633   &        1627   &        1627   &        1554   &        1633   &        1554   \\
\(R^2\)                  &       0.981   &       0.981   &       0.977   &       0.976   &       0.981   &       0.963   \\\hline\hline
\end{tabular}
\justify{In this table, we regress in standard fixed effects regressions several BIS cross-border variables (scaled by deposits) on the principal components extracted from \(\hat{u}_{i,t}\) in our baseline regression and their interactions with the fixed exchange rate dummy. Column (1) uses total cross-border liabilities, columns (2) and (3) break them down into foreign-currency and domestic-currency liabilities and columns (4)-(6) into liabilities provided by banks, non-bank financials and all other sectors of the economy. }
\caption{BIS Variables Regressed on Principal Components: \\ The Role of a Fixed Exchange Rate}
\label{tab:BIS_Interaction}
\end{table}

\clearpage

\begin{figure}
    \centering
    \includegraphics[width=0.75\linewidth]{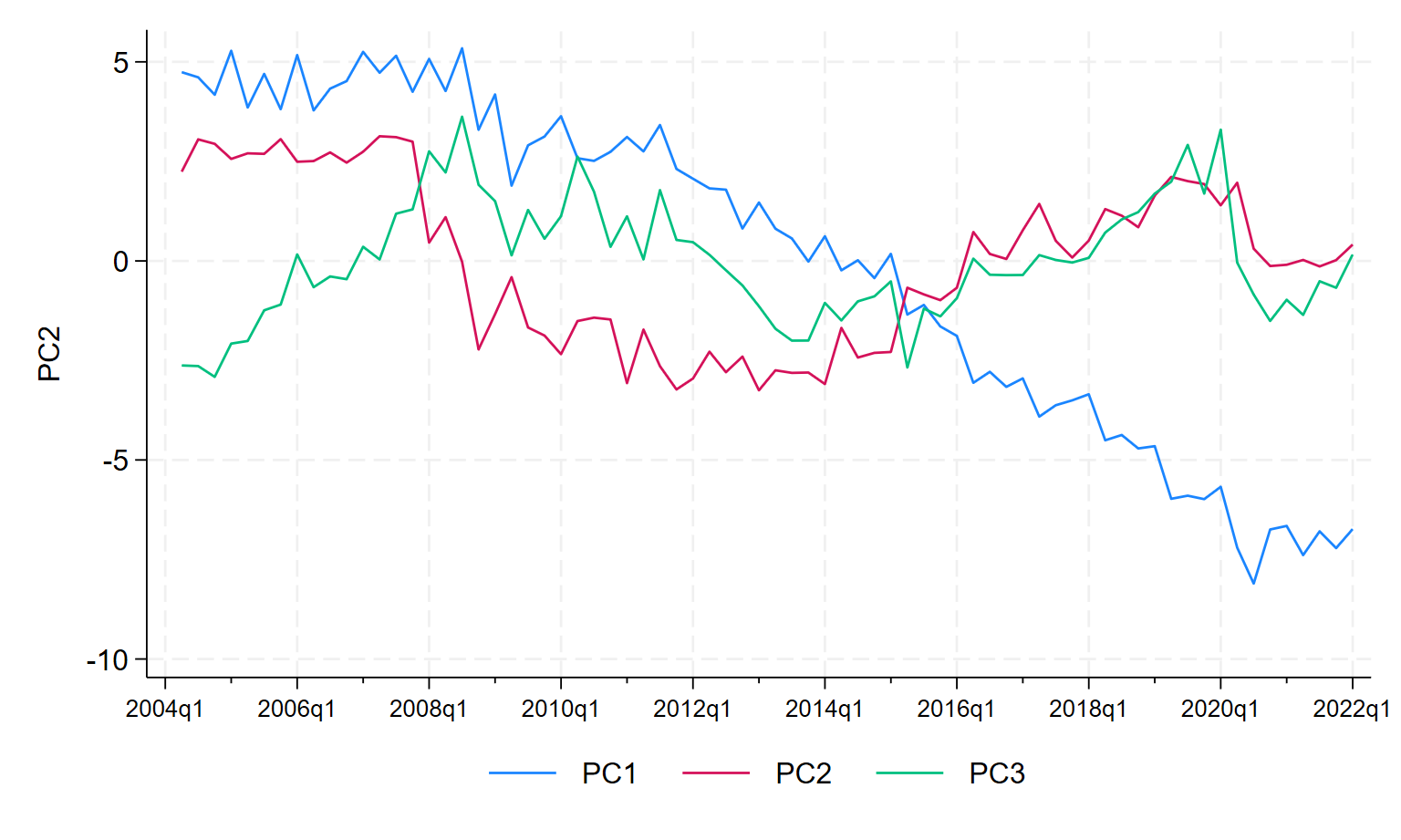}
    \caption{Dynamics of All Three PCs}
    \label{pcfig}
\end{figure}

\begin{figure}
\centering
\includegraphics[width=0.75\linewidth]{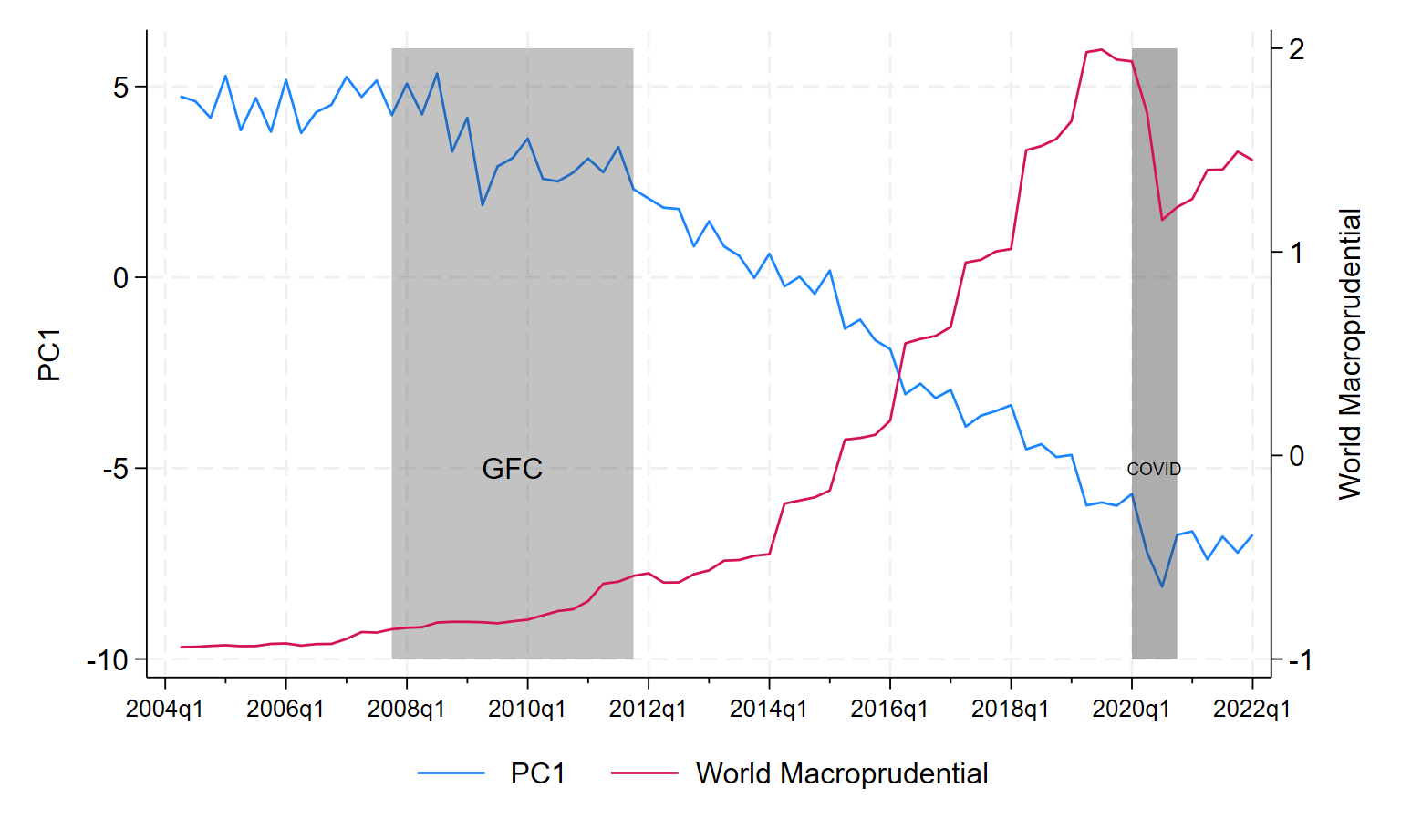}
\caption{PC1 and the World Macroprudential}
\label{fig:pc1}
\end{figure}

\begin{figure}
\centering
\includegraphics[width=0.75\linewidth]{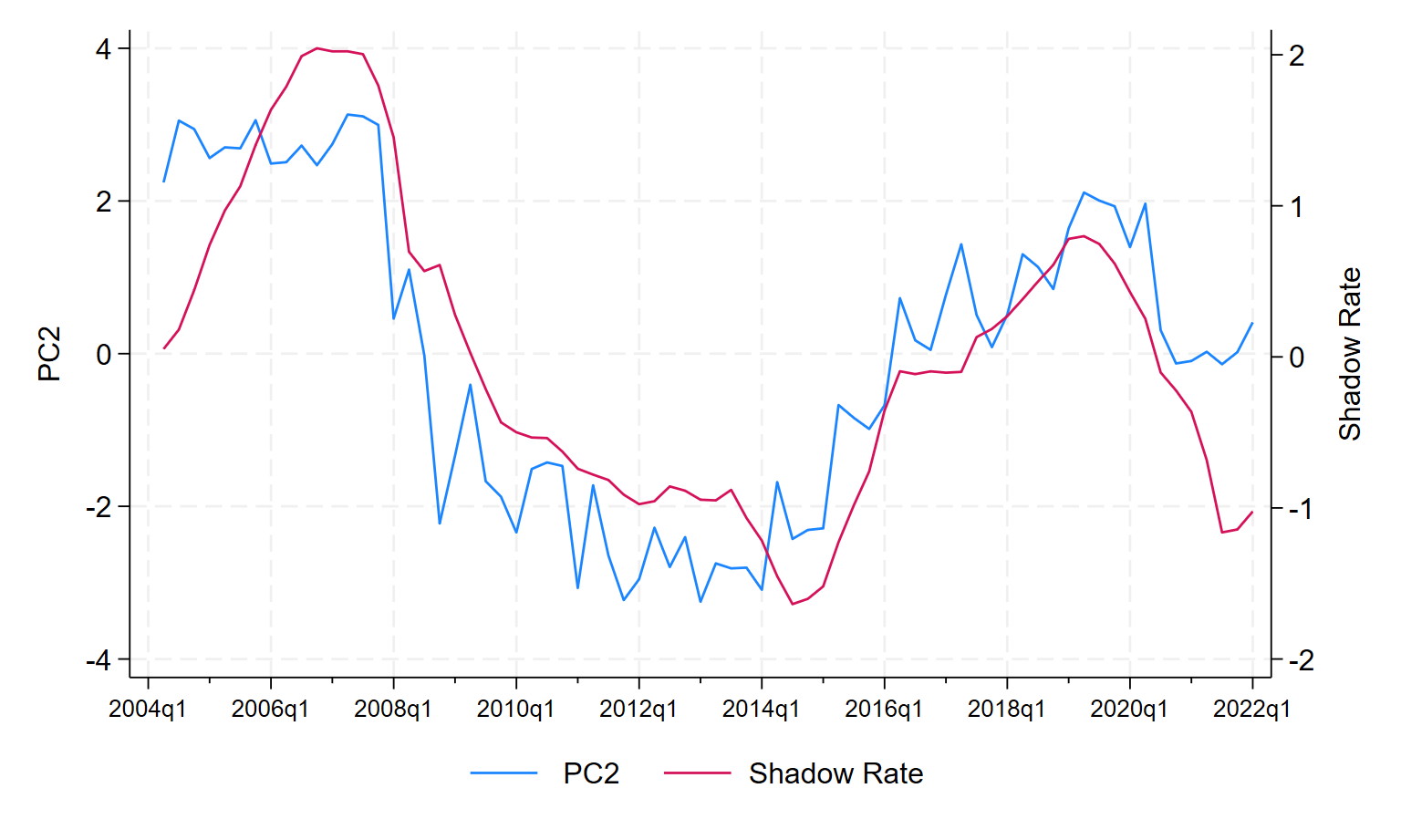}
\caption{PC2 and the US Shadow Rate}
\label{fig:pc2}
\end{figure}

\begin{figure}
\centering
\includegraphics[width=0.75\linewidth]{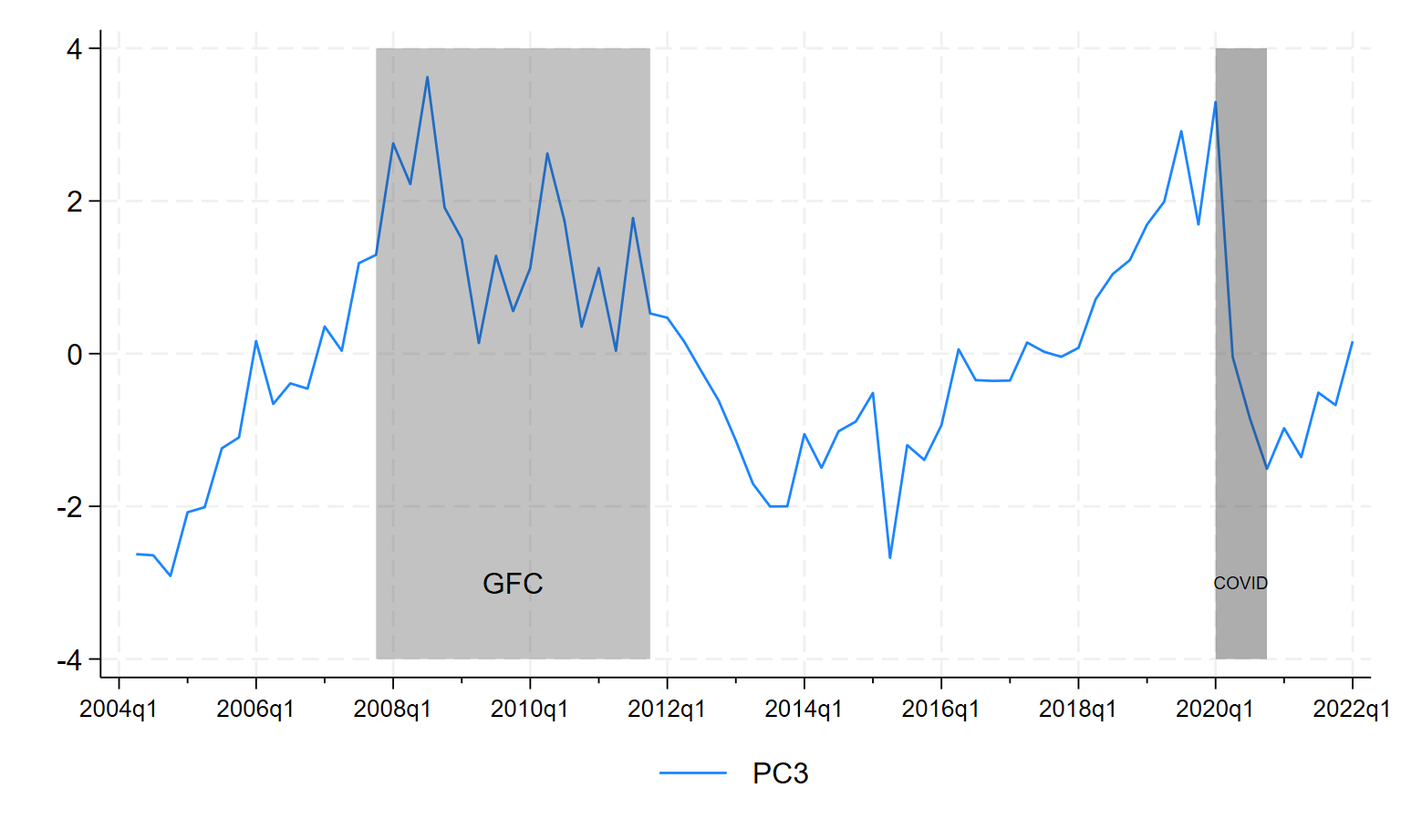}
\caption{PC3 and the Occurrence of Crises}
\label{fig:pc3}
\end{figure}

\end{document}